# COMIC: An Unsupervised Change Detection Method for Heterogeneous Remote Sensing Images Based on Copula Mixtures and Cycle-Consistent Adversarial Networks

Chengxi Li[a], Gang Li[a,*], Zhuoyue Wang[a], Xueqian Wang[a], Pramod K. Varshney[b]

[a]*Tsinghua University, Beijing, 100084, China*
[b]*Syracuse University, Syracuse, NY 13244, USA*

**Abstract**

In this paper, we consider the problem of change detection (CD) with two heterogeneous remote sensing (RS) images. For this problem, an unsupervised change detection method has been proposed recently based on the image translation technique of Cycle-Consistent Adversarial Networks (CycleGANs), where one image is translated from its original modality to the modality of the other image so that the difference map can be obtained by performing arithmetical subtraction. However, the difference map derived from subtraction is susceptible to image translation errors, in which case the changed area and the unchanged area are less distinguishable. To overcome the above shortcoming, we propose a new unsupervised copula mixture and CycleGAN-based CD method (COMIC), which combines the advantages of copula mixtures on statistical modeling and the advantages of CycleGANs on data mining. In COMIC, the pre-event image is first translated from its original modality to the post-event image modality. After that, by constructing a copula mixture, the joint distribution of the features from the heterogeneous images can be learnt according to quantitive analysis of the dependence structure based on the translated image and the original pre-event image, which are of the same modality and contain totally the same objects. Then, we model the CD problem as a binary hypothesis testing problem and derive its test statistics based on the constructed copula mixture. Finally, the difference map can be obtained from the test statistics and the binary change map (BCM) is generated by K-means clustering. We perform experiments on real RS datasets, which demonstrate the superiority of COMIC over the state-of-the-art methods.

## 1. Introduction

In remote sensing (RS), the change detection (CD) task can be regarded as a fusion task [75, 76] which aims to identify possible changes in an area of interest by fusing information contained in the multi-temporal RS images acquired before and after a certain event [1, 2, 58]. The application scenarios of CD are abundant, such as natural disaster assessment, urbanization, land management, and so forth [3-5]. Traditional CD methods mainly focus on homogeneous data, namely RS images of the same modality which are acquired by the same type of sensors [6-11, 68-71]. However, in a vast number of practical situations, homogeneous RS data are not always available, which makes traditional CD methods inapplicable [12, 13]. For instance, when rapid response to sudden eruption of natural disasters is required, one should utilize the data from the first and readily available sensors in spite of their potentially different imaging modalities to facilitate rapid CD. In this case, CD methods with heterogeneous data, namely RS images of different modalities which are acquired by different types of sensors with inter-sensor dependence [55], are well worth studying. For example, heterogeneous data can be acquired by diverse optical sensors, using various multi-spectral bands or offering different levels of spatial resolution. Compared with CD using homogeneous data, CD with heterogeneous data faces new challenges. The pre-event image and the post-event image of different modalities lie in totally different domains[1], and they consequently have different statistical distributions [14]. In those cases, a direct comparison of the images via simple algorithmic operations, as done in traditional CD methods, is infeasible to yield the difference map [15].

[*] Corresponding author.
*E-mail address:* gangli@tsinghua.edu.cn.

CD with heterogeneous RS images has been investigated in the literature [16-22], where the existing methods can be categorized into three classes, namely deep learning methods, nonparametric methods and parametric methods. As representatives of the deep learning methods, in [17], convolutional neural networks (CNNs) are employed to learn features in a common space from the heterogeneous image pair and the difference map is calculated from the feature representations. In [74], a deep neural network-based model is proposed to map heterogeneous RS images to a latent space for CD, which mitigates the influence of different imaging modalities. In [18], an automatic model is proposed to detect the changes based on anomaly feature learning in the residual normal space. An unsupervised change detection method based on image translation and subtraction (ITS) is proposed in [19], which consists of three steps. First, it translates one image from its original modality to the modality of the other image. In this way, the CD problem with heterogeneous RS images is transformed into the CD problem with homogeneous RS images. Second, the difference map is calculated based on the pair of homogeneous images via direct subtraction. Third, changes are identified in the difference map by adopting other techniques, i.e., the clustering techniques, to generate the binary change map (BCM). As representatives of the nonparametric methods, in [20, 21], modality-invariant features are extracted from heterogeneous images and similarity measures are computed based on those features to find the changes. In [73], the heterogeneous RS images are projected to a common feature space, where the images share the same properties, so that CD methods with homogeneous RS images can be applied. In the category of parametric methods, in [22], modelling the dependence structure between the heterogeneous RS images by means of copulas, the conditional copula-based CD technique (C3D) estimates the local statistics in one image when the acquisition conditions are consistent with those of the other image, and derives a similarity measure based on Kullback-Leibler divergence. In [16], physical properties of the sensor are exploited and the joint distribution of the heterogeneous data is inferred through manifold learning. Based on that, the similarity measure is derived to find the changes. In comparison with deep learning methods and non-parametric methods, parametric methods are able to provide analytical guidance based on theoretical analysis, which makes them easily interpretable. On account of the benefits, parametric methods should be adopted when possible, and this category is the main focus of this paper.

Among the existing methods, the advantage of ITS in [19] lies in providing a generalized framework of CD which is totally unsupervised. In other words, ITS does not require any training sample from the unchanged area which is not always easily available in practice. Different from this, the advantage of C3D [22] lies in being able to capture any type of dependence between the heterogeneous images leveraging copulas, which enables performing CD in an interpretable way. However, ITS and C3D both have obvious limitations. The interpretability of ITS needs to be improved[2], and it is very sensitive to image translation errors in the unchanged area based on the fact that its difference map is derived from a simple subtraction operation on the translated image and the original image of the same modality. An individual copula function adopted by C3D is not flexible enough to characterize various dependence structures. Besides, in C3D, an appropriate copula function is selected via visual inspection which may lead to copula misspecification and contribute to performance degradation. Finally, training samples are manually selected from the unchanged area to estimate the copula parameters in C3D, but this is not practical when we are devoid of any knowledge about the locations where changes could possibly occur.

---

[1] In this paper, we have used the terms “images in different domains” and “images of different modalities” interchangeably to describe RS images acquired by different sensors.

[2] The interpretability indicates to what degree humans are able to understand the reasoning behind the results of the method. A more interpretable method is trustier and more comprehensible for humans.

Since copula theory was proposed to characterize the joint distribution of multiple random variables [23, 24], copulas have been widely applied in various financial applications in the past decades [25-30] and have not begun to be used in RS applications until recently [31-33, 57]. In early work, various individual copula functions are constructed to model diverse dependence structures, such as Gaussian copula, Clayton copula, Gumbel copula and so forth [23, 24]. Later, to further enhance the flexibility of modeling asymmetric dependence structures, copula mixtures have been proposed, which decouple the shape of dependence and the degree of dependence [34]. To be more specific, in a copula mixture, the shape of dependence is embodied in the weight parameters while the degree of dependence is embodied in the parameters of individual copulas as the components of the copula mixture [34]. Due to their benefits, copula mixtures have been used to analyze different dependence structures in financial applications such as the Return-volatility dependence in the oil markets and the dependence between the leading stock markets and the emerging markets [34-36]. A copula mixture can be constructed from any sets of copulas, and one should choose or propose an appropriate copula mixture for a specific purpose. For instance, in [34], to model the cases where two markets are likely to crash or boom together, a copula mixture is constructed from a Gaussian copula, a Gumbel copula and a Gumbel survival copula. However, to the best of our knowledge, neither the construction nor the application of copula mixtures has been considered when dealing with CD problems under RS scenarios.

In this paper, to overcome the shortcomings of ITS [19] and C3D [22], we consider the CD problem with heterogeneous RS images and propose a new unsupervised copula mixture and CycleGAN-based CD method (COMIC), which leverages the advantages of copula mixtures and CycleGANs simultaneously. In COMIC, we first translate the pre-event image from its original modality to the post-event image modality using CycleGANs. With the translated image and the original pre-event image of two different modalities, which contain the same objects, we construct a copula mixture to model the joint distribution of the features from heterogeneous images, which captures the dependence structure regarding the association and the tail dependence between them, and we estimate the parameters of the copula mixture via Expectation-Maximization (EM) Algorithm. Subsequently, we model the CD problem as a binary hypothesis testing problem and derive the test statistics based on the constructed copula mixture. After that, the difference map is calculated from the test statistics and the BCM is obtained by K-means clustering. In the end, we run experiments on real RS datasets to validate the superiority of COMIC over the state-of-the-art methods.

This work is motived by the prior work in [19, 22]. However, there are significant differences between COMIC and the methods proposed therein, i.e., ITS and C3D, which are detailed as follows.

1) Both ITS and COMIC employ CycleGANs for image translation, but their purposes of image translation are different and the translated images are utilized in different ways. ITS performs subtraction on the translated image and the original post-event image to obtain the difference map. The images are of the same modality and contain possible changes. In contrast, COMIC utilizes the translated image and the original pre-event image, which are of two different modalities and do not contain any change, to learn the dependence structure and the joint distribution between the heterogeneous data. It can be seen that the proposed method falls into the category of parametric methods and it builds on solid theoretical analysis of the statistical properties of the RS data, which makes it distinct from the deep learning methods in the literature such as ITS.

2) Both C3D and COMIC employ copulas to capture the joint distribution between heterogeneous data. However, copulas are leveraged in two different ways. In C3D, the local statistics of one image are estimated under the same acquisition conditions of the other RS image and the similarity measure based on Kullback-Leibler divergence is computed to highlight the changes. In COMIC, the CD problem is modeled as a binary hypothesis testing problem and the corresponding test statistics derived from the constructed copula mixture are utilized to identify the changes. Besides, COMIC constructs a copula mixture to model the joint distribution of heterogeneous RS data, which is much more flexible than an individual copula function adopted by C3D. Lastly, C3D selects the copula function through visual inspection, which may lead to copula misspecification and performance degradation. In contrast, the copula mixture adopted by COMIC is constructed based on quantitative analysis of the association and the tail dependence between the heterogeneous data, which enhances the CD performance.

Based on their differences, the advantages of COMIC compared to ITS and C3D are listed as follows.

1) By virtue of statistical characterization of the dependence structure between the heterogeneous RS data, COMIC improves the interpretability compared with ITS.

2) ITS is implemented based on the assumption that the translated image and the original image of the same modality have very similar pixel intensities in the unchanged area, which may easily suffer from image translation errors. In contrast, COMIC learns the joint distribution of the heterogeneous data from the original pre-event image and the translated image and it tries to separate the changed area and the unchanged area by distinguishing between two different joint distributions correspondingly. The randomness characterized by the joint distribution implies a better tolerance to image translation errors.

3) COMIC constructs the copula mixture based on quantitive analysis of the statistical dependence structure between the heterogeneous RS data, which is more reliable than the copula selection strategy adopted by C3D via visual inspection. The copula mixture in COMIC is more flexible to characterize various dependence structures, which contributes to better CD performance.

4) COMIC is totally unsupervised, which is more practical than C3D requiring training samples to estimate the copula parameters.

The main contributions in this paper are stated as follows.

1) With the original pre-event image and the image translated from the pre-event image modality to the post-event image modality via CycleGANs, we construct a copula mixture according to quantitive analysis of statistical dependence between the heterogeneous RS images and we estimate the parameters of the constructed copula mixture using EM algorithm.

2) We model the CD problem as a binary hypothesis testing problem and derive its test statistics based on the constructed copula mixture, which yields the difference map and the BCM.

3) We conduct a number of experiments on real RS datasets, which demonstrate that COMIC outperforms the state-of-the-art methods in terms of the CD performance.

The organization of the rest of this paper is as follows. In Section 2, the considered problem is formally formulated. In Section 3, we present the steps of COMIC. In Section 4, we run experiments and show the experimental results to demonstrate the superiority of COMIC compared with the state-of-the-art methods. Finally, we draw our conclusions in Section 5.

## 2. Problem formulation

We consider two registered images acquired by two sensors of disparate types, which include a pre-event image $\mathbf{X} \sim \mathbb{R}^{M \times N \times C_X}$ and a post-event image $\mathbf{Y} \sim \mathbb{R}^{M \times N \times C_Y}$ [16-22]. The height and the width of the images are denoted by *M* and *N*. The channel numbers in $\mathbf{X}$ and $\mathbf{Y}$ are denoted by $C_X$ and $C_Y$, respectively, where $C_X > 1$ or $C_Y > 1$ implies that there are multiple channels in an image. For instance, multispectral images are composed of multiple bands [37] and polarimetric synthetic aperture radar images have a number of polarimetric channels [45]. The pixel intensities in $\mathbf{X}$ and $\mathbf{Y}$ are denoted by $x(m,n,c)$ and $y(m,n,c)$, respectively, $m=1,\ldots,M$, $n=1,\ldots,N$, $c=1,\ldots,C_X$ or $c=1,\ldots,C_Y$. The aim of CD is to generate a BCM denoted by $\mathbf{B}\sim\mathbb{N}^{M\times N}$ to highlight the changes between $\mathbf{X}$ and $\mathbf{Y}$ [16-22], where the $(m,n)$-th element is denoted by $b(m,n)$. In $\mathbf{B}$, $b(m,n)=1$ indicates that pixel $(m,n)$ is in the changed area, while $b(m,n)=0$ implies the opposite.

## 3. The proposed method: COMIC

In this section, we introduce the proposed method, which consists of six steps: 1) Image translation; 2) Feature extraction; 3) Copula mixture construction; 4) Parameter estimation; 5) Test statistic derivation; 6) Difference map and BCM calculation. The flowchart of COMIC is presented in Fig. 1.

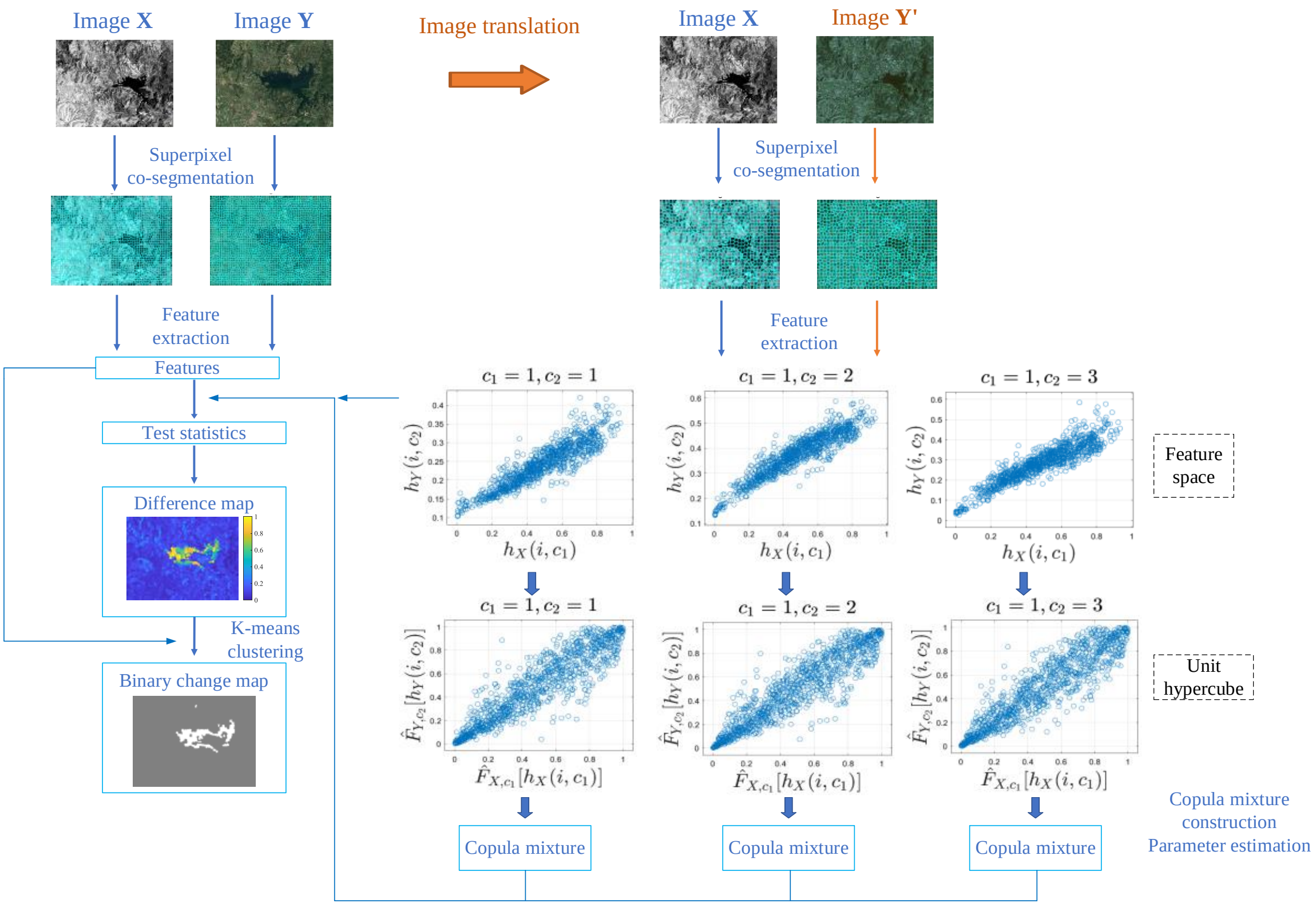

Fig. 1. The flowchart of COMIC. It consists of six steps: (a) The pre-event image $\mathbf{X}$ is translated into image $\mathbf{Y}'$ which is of the same modality as the original post-event image $\mathbf{Y}$. (b) Features are extracted from the translated image and the original pre-event image of two different modalities. (c) A copula mixture is constructed to model the joint distribution of the features from heterogeneous images. (d) The parameters of the copula mixture are estimated via EM Algorithm. (e) The CD problem is modeled as a binary hypothesis testing problem and the test statistics are derived based on the constructed copula mixture. (f) The difference map is calculated from the test statistics and the BCM is obtained by K-means clustering.

*3.1. Image translation*

Image translation aims to translate one image from its original modality to another modality by learning a mapping between the source domain and the target domain. As the first step of the proposed method, we adopt image translation techniques to translate the pre-event image from its original modality to the post-event image modality. Note that the translated image and the original pre-event image, which are of two different modalities, contain the same objects. Those two images enable learning the dependence structure and the joint distribution of features extracted from the heterogeneous RS images in the subsequent steps of COMIC.

Traditional image translation techniques, such as Pix2pix with Conditional Adversarial Networks (cGANs) [38], require paired training data to learn the mapping. However, under the practical scenarios of CD, collecting paired training data is often unfeasible or at a fairly high cost. To overcome this shortcoming, in [39], an image translation technique with CycleGANs is proposed, which is able to translate images between two different domains in the absence of paired training data. Due to its benefits, in this work, the image translation technique with CycleGANs is utilized.

As done in [19], we first collect unpaired training samples by applying an $M_1 \times N_1$ sliding window of step size $\lambda$ on images $\mathbf{X}$ and $\mathbf{Y}$, which are denoted by $\left\{\tilde{\mathbf{X}}_t \sim \mathbb{R}^{M_1 \times N_1 \times C_X}, t=1,...,T\right\}$ and $\left\{\tilde{\mathbf{Y}}_t \sim \mathbb{R}^{M_1 \times N_1 \times C_Y}, t=1,...,T\right\}$. The number of training samples collected from each image is $T$. Let us define two mappings, i.e., $\mathcal{G}_{X\to Y}$ and $\mathcal{G}_{Y\to X}$, and two discriminators, i.e., $\mathcal{D}_X$ and $\mathcal{D}_Y$. As shown in Fig. 2, the mapping $\mathcal{G}_{X\to Y}$ aims to translate a patch $\tilde{\mathbf{X}}$ from the pre-event image modality to the post-event image modality, and the mapping $\mathcal{G}_{Y\to X}$ aims to translate a patch $\tilde{\mathbf{Y}}$ from the post-event image modality to the pre-event image modality. $\mathcal{D}_X$ and $\mathcal{D}_Y$ are two adversarial discriminators, where the former aims to discriminate between $\tilde{\mathbf{X}}$ and the translated patch $\mathcal{G}_{Y\to X}\left(\tilde{\mathbf{Y}}\right)$, and the latter aims to distinguish between $\tilde{\mathbf{Y}}$ and $\mathcal{G}_{X\to Y}\left(\tilde{\mathbf{X}}\right)$. With unpaired samples $\left\{\tilde{\mathbf{X}}_t, t=1,...,T\right\}$ and $\left\{\tilde{\mathbf{Y}}_t, t=1,...,T\right\}$, the mappings $\mathcal{G}_{X\to Y}$ and $\mathcal{G}_{Y\to X}$ can be trained by solving the following problem [39]:

$$\mathcal{G}_{X\to Y}, \mathcal{G}_{Y\to X} = \arg \min_{\mathcal{G}_{X\to Y}, \mathcal{G}_{Y\to X}} \max_{\mathcal{D}_X, \mathcal{D}_Y} \mathcal{L}_{\text{overall}}, \tag{1}$$

where $\mathcal{L}_{\text{overall}}$ is the loss function defined as

$$\mathcal{L}_{\text{overall}} = \mathcal{L}_{\text{GAN}}\left(\mathcal{G}_{X\to Y}, \mathcal{D}_Y\right) + \mathcal{L}_{\text{GAN}}\left(\mathcal{G}_{Y\to X}, \mathcal{D}_X\right) + \beta \mathcal{L}_{\text{cyc}}\left(\mathcal{G}_{X\to Y}, \mathcal{G}_{Y\to X}\right), \tag{2}$$

with

$$\mathcal{L}_{\text{GAN}}\left(\mathcal{G}_{X\to Y}, \mathcal{D}_Y\right) = \mathbb{E}_{\tilde{\mathbf{Y}}\sim p_{\text{data}}(\tilde{\mathbf{Y}})}\left[\log \mathcal{D}_Y\left(\tilde{\mathbf{Y}}\right)\right] + \mathbb{E}_{\tilde{\mathbf{X}}\sim p_{\text{data}}(\tilde{\mathbf{X}})}\left[\log\left(1 - \mathcal{D}_Y\left(\mathcal{G}_{X\to Y}\left(\tilde{\mathbf{X}}\right)\right)\right)\right], \tag{3}$$

$$\mathcal{L}_{\text{GAN}}\left(\mathcal{G}_{Y\to X}, \mathcal{D}_X\right) = \mathbb{E}_{\tilde{\mathbf{X}}\sim p_{\text{data}}(\tilde{\mathbf{X}})}\left[\log \mathcal{D}_X\left(\tilde{\mathbf{X}}\right)\right] + \mathbb{E}_{\tilde{\mathbf{Y}}\sim p_{\text{data}}(\tilde{\mathbf{Y}})}\left[\log\left(1 - \mathcal{D}_X\left(\mathcal{G}_{Y\to X}\left(\tilde{\mathbf{Y}}\right)\right)\right)\right], \tag{4}$$

$$\mathcal{L}_{\text{cyc}}\left(\mathcal{G}_{X\to Y}, \mathcal{G}_{Y\to X}\right) = \mathbb{E}_{\tilde{\mathbf{X}}\sim p_{\text{data}}(\tilde{\mathbf{X}})}\left[\left\|\mathcal{G}_{Y\to X}\left(\mathcal{G}_{X\to Y}\left(\tilde{\mathbf{X}}\right)\right) - \tilde{\mathbf{X}}\right\|_1\right] + \mathbb{E}_{\tilde{\mathbf{Y}}\sim p_{\text{data}}(\tilde{\mathbf{Y}})}\left[\left\|\mathcal{G}_{X\to Y}\left(\mathcal{G}_{Y\to X}\left(\tilde{\mathbf{Y}}\right)\right) - \tilde{\mathbf{Y}}\right\|_1\right], \tag{5}$$

and $\beta$ is the controlling factor of the relative importance of different objectives. In (3)-(5), $p_{\text{data}}\left(\tilde{\mathbf{X}}\right)$ and $p_{\text{data}}\left(\tilde{\mathbf{Y}}\right)$ represent the statistical distributions of patches from the pre-event image modality and from the post-event image modality, respectively, $\mathbb{E}\left(\bullet\right)$ denotes the statistical expectation, and $\left\|\bullet\right\|_1$ is the $L_1$ norm.

The loss function in (2) indicates that the objective of training the mappings is two-fold. First, the adversarial losses in (3) and (4) ensure that the distribution of the generated patches matches that of the target modality. Second, the cycle consistency loss in (5) ensures that the contextual information in the generated patches is consistent with that in the original image of the source modality. When the training is completed, with the mappings $\mathcal{G}_{X\to Y}$ and $\mathcal{G}_{Y\to X}$, we can translate image **X** into image $\mathbf{Y}'$ which is of the same modality as the original post-event image **Y**. To this end, image **X** is cut into small patches of size $M_1 \times N_1 \times C_X$, which are subsequently fed into $\mathcal{G}_{X\to Y}$. The output patches of the mapping $\mathcal{G}_{X\to Y}$ are stitched together to yield $\mathbf{Y}' \sim \mathbb{R}^{M\times N\times C_Y}$.

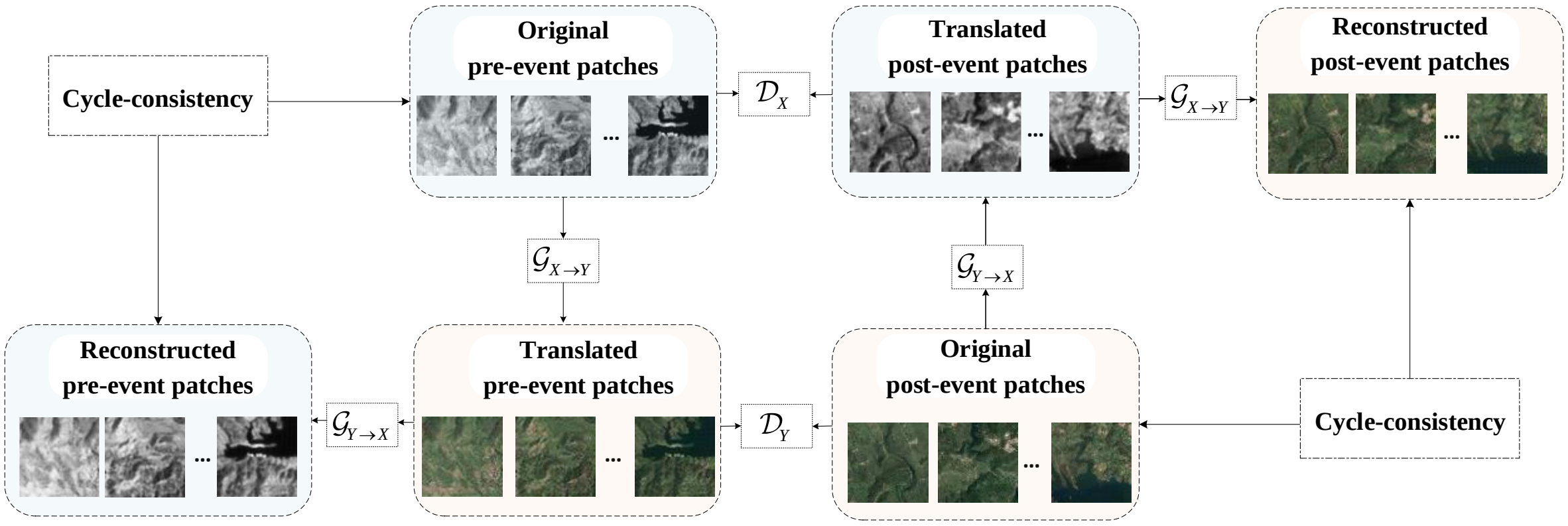

Fig. 2. The mappings and discriminators in the image translation model.

*3.2. Feature extraction*

Note that the objects contained in the original pre-event image **X** and the translated image $\mathbf{Y}'$ are totally the same and the difference between the two images lies in their imaging modalities. To be more specific, image **X** is of the pre-event image modality, while image $\mathbf{Y}'$ is of the post-event image modality. In Section 3.2, we show how to extract features from the two images.

Superpixels are atomic regions of pixel groups that are perceptually meaningful [46, 47, 56]. Performing feature extraction at the superpixel level is convenient and reduces the complexity of subsequent processing [48]. Based on their advantages, superpixels have been widely adopted as the basic processing units in various RS applications such as vessel detection [40], band selection [41], and image segmentation [42]. Likewise, in COMIC, based on the translated image and the original pre-event image, features are extracted at the superpixel level following the steps below. First, we perform superpixel co-segmentation[3] as done in [20]. To be more specific, image **X** and $\mathbf{Y}'$ are segmented into superpixels via simple liner iterative clustering (SLIC), respectively, and then two individual segmentation maps are merged into a superpixel co-segmentation map. In this case, each superpixel in the co-segmentation map has a homogeneous interior, both in **X** and $\mathbf{Y}'$. Let us denote the co-segmentation map as $\mathbf{S} \in \mathbb{N}^{M\times N}$ with $N_S$ superpixels, whose elements are represented by $s(m,n)$, $m=1,...,M$, $n=1,...,N$. In the co-segmentation map, if pixel $(m,n)$ is in the $i$-th superpixel, we have $s(m,n)=i$, $i=1,2,...,N_S$. Second, we extract features from the images based on the co-segmentation map $\mathbf{S}$ by calculating the mean of the pixel intensities in each superpixel. To be more concrete, for image **X**,

[3] The code is available at https://github.com/yulisun/IRG-McS.

the feature vector of the $i$-th superpixel is obtained as

$$\mathbf{h}_X(i)=\left[h_X(i,1),...,h_X(i,C_X)\right], i=1,2,...,N_S, \tag{6}$$

where

$$h_X(i,c_1)=\frac{\sum_{m,n} I\left[s(m,n),i\right]x(m,n,c_1)}{\sum_{m,n} I\left[s(m,n),i\right]}, c_1=1,...,C_X, i=1,...,N_S. \tag{7}$$

In (7), we have

$$I(a,b)=\begin{cases}1, & \text{if } a=b,\\ 0, & \text{otherwise.}\end{cases} \tag{8}$$

Likewise, for image $\mathbf{Y}'$, we can express the feature vector of the $i$-th superpixel as

$$\mathbf{h}_Y(i)=\left[h_Y(i,1),...,h_Y(i,C_Y)\right], i=1,2,...,N_S, \tag{9}$$

where

$$h_Y(i,c_2)=\frac{\sum_{m,n} I\left[s(m,n),i\right]y(m,n,c_2)}{\sum_{m,n} I\left[s(m,n),i\right]}, c_2=1,...,C_Y, i=1,...,N_S. \tag{10}$$

### *3.3. Copula mixture construction*

With the features extracted from the heterogeneous RS images as shown in Section 3.2, we elaborate how to construct a copula mixture according to quantitive analysis of the dependence structure to capture the joint distribution of the features.

First, we introduce the quantitive measures of the dependence structure adopted in this paper, which include the measure of association and the measure of tail dependence [49]. Different from correlation which describes the linear dependence, association captures dependence in a more general sense, which refers to dependence of any possible type and indicates to what degree a larger value of one variable is associated with a larger value of the other [50]. Kendall's $\tau$ is defined as an important measure of association of two random variables $(h_1,h_2)$, which is expressed as [49]

$$\tau(h_1,h_2)\triangleq \Pr\left[(h_{11}-h_{21})(h_{12}-h_{22})>0\right]-\Pr\left[(h_{11}-h_{21})(h_{12}-h_{22})<0\right]. \tag{11}$$

In (11), $(h_{11},h_{12})$ and $(h_{21},h_{22})$ are two independent sets of random variables with the same distribution as $(h_1,h_2)$. A positive value of $\tau(h_1,h_2)$ implies that the random variables $(h_1,h_2)$ are positively associated, while a negative value implies the opposite [49]. Kendall's $\tau$ of the features extracted from the heterogeneous RS data can be estimated in the following way [51]:

$$\tau_{XY,c_1,c_2}=\frac{2\sum_{i=1}^{N_S-1}\sum_{j=i+1}^{N_S}\xi\left[h_X(i,c_1),h_X(j,c_1),h_Y(i,c_2),h_Y(j,c_2)\right]}{N_S(N_S-1)}, \forall c_1, \forall c_2, \tag{12}$$

where

$$\xi\left[h_X(i,c_1),h_X(j,c_1),h_Y(i,c_2),h_Y(j,c_2)\right]=\begin{cases}1, & \text{if } \left[h_X(i,c_1)-h_X(j,c_1)\right]\left[h_Y(i,c_2)-h_Y(j,c_2)\right]>0,\\ 0, & \text{if } \left[h_X(i,c_1)-h_X(j,c_1)\right]\left[h_Y(i,c_2)-h_Y(j,c_2)\right]=0,\\ -1, & \text{if } \left[h_X(i,c_1)-h_X(j,c_1)\right]\left[h_Y(i,c_2)-h_Y(j,c_2)\right]<0.\end{cases} \tag{13}$$

Kendall's $\tau$ characterizes the dependence in the whole space where the random variables take values. In contrast, tail dependence focuses on the dependence in the upper-right and lower-left quadrants in the whole space [49]. To be more specific, the dependence in the upper-right quadrant is referred to as the upper tail dependence while the dependence in the lower-left quadrant is referred to as the lower tail dependence. The measures of the lower tail dependence and the upper tail dependence of two random variables $(h_1, h_2)$ are defined as [49]

$$\eta^{\text{lower}} \triangleq \lim_{t \to 0^+} \Pr\left[h_2 < F_2^{-1}(t) \middle| h_1 < F_1^{-1}(t)\right], \tag{14}$$

and

$$\eta^{\text{upper}} \triangleq \lim_{t \to 1^-} \Pr\left[h_2 > F_2^{-1}(t) \middle| h_1 > F_1^{-1}(t)\right], \tag{15}$$

respectively, where $F_1^{-1}$ and $F_2^{-1}$ are the inverse cumulative distribution functions (cdfs) of $h_1$ and $h_2$.

Next, we show how to construct a copula mixture to capture the joint distribution of the features from heterogeneous images based on the above quantitive measures. Let us denote the joint cdf of $h_X(i, c_1)$ and $h_Y(i, c_2)$ by $F_{XY,c_1,c_2}\left[h_X(i,c_1), h_Y(i,c_2)\right]$, with channel $c_1$ in image $\mathbf{X}$ and channel $c_2$ in the translated image $\mathbf{Y}'$. According to Sklar's Theorem [24], there exists a unique copula function $F_{\text{Copula},c_1,c_2}$ which satisfies

$$F_{XY,c_1,c_2}\left[h_X(i,c_1), h_Y(i,c_2)\right] = F_{\text{Copula},c_1,c_2}\left\{F_{X,c_1}\left[h_X(i,c_1)\right], F_{Y,c_2}\left[h_Y(i,c_2)\right]\right\}, \forall i. \tag{16}$$

$c_1 = 1,\ldots,C_X$, $c_2 = 1,\ldots,C_Y$. In (16), the marginal cdfs of $h_X(i,c_1)$ and $h_Y(i,c_2)$ are denoted by $F_{X,c_1}$ and $F_{Y,c_2}$, and the empirical versions of them, i.e., $\hat{F}_{X,c_1}$ and $\hat{F}_{Y,c_2}$, can be obtained in the following way [52]:

$$\hat{F}_{X,c_1}(h) \approx \frac{1}{N_S}\sum_{i=1}^{N_S} \chi\left[h - h_X(i,c_1)\right], \forall c_1, \tag{17}$$

$$\hat{F}_{Y,c_2}(h) \approx \frac{1}{N_S}\sum_{i=1}^{N_S} \chi\left[h - h_Y(i,c_2)\right], \forall c_2, \tag{18}$$

where $\chi(\bullet)$ denotes the step function. Note that the marginal cdfs could also be estimated parametrically, but in COMIC we estimate the marginal cdfs nonparametrically as in (17) and (18), which is free of the error of marginal specifications [34]. Assuming that the second partial derivatives of $F_{\text{Copula},c_1,c_2}$ exist, the copula density function corresponding to $F_{\text{Copula},c_1,c_2}$ is given as [24]

$$f_{\text{Copula},c_1,c_2}(u_1,u_2) = \frac{\partial^2 F_{\text{Copula},c_1,c_2}(u_1,u_2)}{\partial u_1 \partial u_2}. \tag{19}$$

Based on (16)-(19), our aim of constructing a copula function to capture the joint distribution of the features from heterogeneous images can be equivalently expressed as constructing the copula density function $f_{\text{Copula},c_1,c_2}$ in order to fit the data $\left\{\hat{F}_{X,c_1}\left[h_X(i,c_1)\right], \hat{F}_{Y,c_2}\left[h_Y(i,c_2)\right], \forall i\right\}$, $\forall c_1$, $\forall c_2$. Note that heterogeneous images have different imaging modalities, and the intensities in the two images corresponding to the same object rely on different kinds of physical properties. In other words, the intensities in the same location may not be positively associated [43, 44]. Constructing a copula which is flexible enough for both the positive association and the negative association is more complex than constructing a copula which only captures positive association. From this perspective, to reduce the complexity of copula construction, transformation is performed on $\left\{\hat{F}_{X,c_1}\left[h_X(i,c_1)\right], \hat{F}_{Y,c_2}\left[h_Y(i,c_2)\right], \forall i\right\}$ based on the value of $\tau_{XY,c_1,c_2}$ according to the following rule:

$$\breve{F}_{Y,c_1,c_2}\left[h_Y\left(i,c_2\right)\right]=\begin{cases}1-\hat{F}_{Y,c_2}\left[h_Y\left(i,c_2\right)\right], & \text{if } \tau_{XY,c_1,c_2}<0,\\ \hat{F}_{Y,c_2}\left[h_Y\left(i,c_2\right)\right], & \text{otherwise,}\end{cases}\ \forall i,\forall c_1,\forall c_2, \tag{20}$$

where $\hat{F}_{X,c_1}$ and $\hat{F}_{Y,c_2}$ are given as (17) and (18). Based on (20), it can be seen that $\hat{F}_{X,c_1}\left[h_X\left(i,c_1\right)\right]$ and $\breve{F}_{Y,c_1,c_2}\left[h_Y\left(i,c_2\right)\right]$ are always positively associated, no matter $\hat{F}_{X,c_1}\left[h_X\left(i,c_1\right)\right]$ and $\hat{F}_{Y,c_2}\left[h_Y\left(i,c_2\right)\right]$ are positively associated or negatively associated, $\forall c_1$, $\forall c_2$. Let us denote the copula density function of $\left\{\hat{F}_{X,c_1}\left[h_X\left(i,c_1\right)\right],\breve{F}_{Y,c_1,c_2}\left[h_Y\left(i,c_2\right)\right],\forall i\right\}$ as $\breve{f}_{\text{Copula},c_1,c_2}$. From (20), the copula density function $f_{\text{Copula},c_1,c_2}$ of $\left\{\hat{F}_{X,c_1}\left[h_X\left(i,c_1\right)\right],\hat{F}_{Y,c_2}\left[h_Y\left(i,c_2\right)\right],\forall i\right\}$ and the copula density function $\breve{f}_{\text{Copula},c_1,c_2}$ are related as follows:

$$\breve{f}_{\text{Copula},c_1,c_2}\left(u_1,u_2\right)=\begin{cases}f_{\text{Copula},c_1,c_2}\left(u_1,1-u_2\right), & \text{if } \tau_{XY,c_1,c_2}<0,\\ f_{\text{Copula},c_1,c_2}\left(u_1,u_2\right), & \text{otherwise,}\end{cases}\ \forall i,\forall c_1,\forall c_2. \tag{21}$$

Based on the above analysis, our aim can be equivalently expressed as constructing a copula density function $\breve{f}_{\text{Copula},c_1,c_2}$ which captures the positive association between $\hat{F}_{X,c_1}\left[h_X\left(i,c_1\right)\right]$ and $\breve{F}_{Y,c_1,c_2}\left[h_Y\left(i,c_2\right)\right]$.

After that, we show how to construct the copula density function $\breve{f}_{\text{Copula},c_1,c_2}$ from $\left\{\hat{F}_{X,c_1}\left[h_X\left(i,c_1\right)\right],\breve{F}_{Y,c_1,c_2}\left[h_Y\left(i,c_2\right)\right],\forall i\right\}$ based on the tail dependence. Primarily, we should bear in mind that the dependence structure may be asymmetrical, which means either the upper tail dependence or the lower tail dependence could be dominant. To identify which one is dominant, we can estimate the measures defined in (14) and (15) for the features extracted from the heterogeneous RS data in the following way [49]:

$$\begin{aligned}\hat{\eta}^{\text{lower}}_{XY,c_1,c_2}&=\frac{1}{\Gamma\left(\sqrt{N_S}\right)}\sum_{i=1}^{N_S}\chi\left(\frac{\Gamma\left(\sqrt{N_S}\right)}{N_S}-\hat{F}^0_{X,c_1}\left[h_X\left(i,c_1\right)\right]\right)\chi\left(\frac{\Gamma\left(\sqrt{N_S}\right)}{N_S}-\breve{F}^0_{Y,c_1,c_2}\left[h_Y\left(i,c_2\right)\right]\right),\\ \hat{\eta}^{\text{upper}}_{XY,c_1,c_2}&=\frac{1}{\Gamma\left(\sqrt{N_S}\right)}\sum_{i=1}^{N_S}\chi\left(\hat{F}^0_{X,c_1}\left[h_X\left(i,c_1\right)\right]-1+\frac{\Gamma\left(\sqrt{N_S}\right)}{N_S}\right)\chi\left(\breve{F}^0_{Y,c_1,c_2}\left[h_Y\left(i,c_2\right)\right]-1+\frac{\Gamma\left(\sqrt{N_S}\right)}{N_S}\right),\end{aligned} \tag{22}$$

where $\Gamma(\bullet)$ represents the round operation toward negative infinity. If $\hat{\eta}^{\text{lower}}_{XY,c_1,c_2}>\hat{\eta}^{\text{upper}}_{XY,c_1,c_2}$, the lower tail dependence is dominant. Otherwise, the upper tail dependence is dominant. To enhance the flexibility of modelling various asymmetrical dependence structures, we construct a copula mixture, which can be expressed as a weighted sum of two individual copulas. One individual copula is the Gaussian copula, which is symmetrical and informative for modelling the overall dependence structure between the variables [24, 34]. The second one is either the Clayton copula or the Clayton survival copula. It is known that both the Clayton copula and the Clayton survival copula model positive asymmetric tail dependence. However, in Clayton copula, lower tail dependence is stronger than upper tail dependence, while in Clayton survival copula upper tail dependence is stronger than lower tail dependence [49]. For the considered problem, when the lower tail dependence is dominant, the second individual copula is selected to be the Clayton copula. When the upper tail dependence is dominant, the second individual copula is selected to be the Clayton survival copula. In this way, the copula function of the copula mixture is constructed as

$$F_{\text{mixture},c_1,c_2}\left(u_1,u_2;\rho_{c_1,c_2},\theta_{c_1,c_2},w_{c_1,c_2}\right)=$$
$$\begin{cases} w_{c_1,c_2}F_{\text{Gaussian}}\left(u_1,u_2;\rho_{c_1,c_2}\right)+\left(1-w_{c_1,c_2}\right)F_{\text{Clayton}}\left(u_1,u_2;\theta_{c_1,c_2}\right), & \text{if } \hat{\eta}^{\text{lower}}_{XY,c_1,c_2}>\hat{\eta}^{\text{upper}}_{XY,c_1,c_2}, \\ w_{c_1,c_2}F_{\text{Gaussian}}\left(u_1,u_2;\rho_{c_1,c_2}\right)+\left(1-w_{c_1,c_2}\right)F_{\text{S-Clayton}}\left(u_1,u_2;\theta_{c_1,c_2}\right), & \text{otherwise,}\end{cases}\quad w_{c_1,c_2}\in[0,1]. \tag{23}$$

By assigning appropriate parameters $\left\{\rho_{c_1,c_2},\theta_{c_1,c_2},w_{c_1,c_2},\forall c_1,\forall c_2\right\}$ to the copula mixture in (23), various patterns of dependence structures can be captured, which include but not limited to those modeled by an individual copula. In (23), $F_{\text{Gaussian}}$ is the Gaussian copula function expressed as [24]

$$F_{\text{Gaussian}}\left(u_1,u_2;\rho\right)=\phi_\rho\left(\phi^{-1}\left(u_1\right),\phi^{-1}\left(u_2\right)\right),\rho\in[-1,1], \tag{24}$$

where $\phi_\rho$ is the cdf of standardized bivariate normal distribution with correlation $\rho$, and $\phi$ is the cdf of standardized univariate normal distribution. $F_{\text{Clayton}}$ is the Clayton copula function expressed as [24]

$$F_{\text{Clayton}}\left(u_1,u_2;\theta\right)=\left(u_1^{-\theta}+u_2^{-\theta}-1\right)^{-\frac{1}{\theta}},\theta\in(0,+\infty), \tag{25}$$

and $F_{\text{S-Clayton}}$ is the Clayton survival copula function given as [24]

$$F_{\text{S-Clayton}}\left(u_1,u_2;\theta\right)=u_1+u_2-1+\left[\left(1-u_1\right)^{-\theta}+\left(1-u_2\right)^{-\theta}-1\right]^{-\frac{1}{\theta}},\theta\in(0,+\infty). \tag{26}$$

It can be seen from (23) that, $\rho_{c_1,c_2}$ and $\theta_{c_1,c_2}$ are the parameters that reflect the degree of dependence and $w_{c_1,c_2}$ reflects the shape of dependence. Correspondingly, the copula density function of the constructed copula mixture can be written as

$$\breve{f}_{\text{Copula},c_1,c_2}\left(u_1,u_2;\rho_{c_1,c_2},\theta_{c_1,c_2},w_{c_1,c_2}\right)=$$
$$\begin{cases} w_{c_1,c_2}f_{\text{Gaussian}}\left(u_1,u_2;\rho_{c_1,c_2}\right)+\left(1-w_{c_1,c_2}\right)f_{\text{Clayton}}\left(u_1,u_2;\theta_{c_1,c_2}\right), & \text{if } \hat{\eta}^{\text{lower}}_{XY,c_1,c_2}>\hat{\eta}^{\text{upper}}_{XY,c_1,c_2}, \\ w_{c_1,c_2}f_{\text{Gaussian}}\left(u_1,u_2;\rho_{c_1,c_2}\right)+\left(1-w_{c_1,c_2}\right)f_{\text{S-Clayton}}\left(u_1,u_2;\theta_{c_1,c_2}\right), & \text{otherwise,}\end{cases}\quad w_{c_1,c_2}\in[0,1], \tag{27}$$

where [24]

$$f_{\text{Gaussian}}\left(u_1,u_2;\rho\right)=\frac{1}{\sqrt{1-\rho^2}}\exp\left\{-\frac{\rho^2\left[\left(\phi^{-1}\left(u_1\right)\right)^2+\left(\phi^{-1}\left(u_2\right)\right)^2\right]-2\rho\phi^{-1}\left(u_1\right)\phi^{-1}\left(u_2\right)}{2\left(1-\rho^2\right)}\right\}, \tag{28}$$

$$f_{\text{Clayton}}\left(u_1,u_2;\theta\right)=\left(1+\theta\right)\left(u_1u_2\right)^{-1-\theta}\left(u_1^{-\theta}+u_2^{-\theta}-1\right)^{-\frac{1}{\theta}-2}, \tag{29}$$

$$f_{\text{S-Clayton}}\left(u_1,u_2;\theta\right)=\left(1+\theta\right)\left[u_1\left(1-u_2\right)\right]^{-1-\theta}\left[u_1^{-\theta}+\left(1-u_2\right)^{-\theta}-1\right]^{-\frac{1}{\theta}-2}. \tag{30}$$

For instance, in Fig. 3, we provide the contour plots of the copula density functions of the constructed copula mixture and the corresponding values on the diagonal $u_1=u_2$. Note that individual copula functions are special cases of the copula mixture, and the copula mixture has a higher degree of flexibility of dependence modelling compared with the individual ones therein.

When $\breve{f}_{\text{Copula},c_1,c_2}$ is successfully constructed as (27), the copula density function $f_{\text{Copula},c_1,c_2}$ can be obtained based on (21). In this case, from (21), taking the second-order derivative on both sides of (16), we can express the joint probability density function (pdf) of $h_X\left(i,c_1\right)$ and $h_Y\left(i,c_2\right)$ as

$$
\begin{aligned}
& f_{XY,c_1,c_2}\left[h_X\left(i,c_1\right),h_Y\left(i,c_2\right)\right] \\
& = \frac{\partial^2}{\partial h_X\left(i,c_1\right)\partial h_Y\left(i,c_2\right)} F_{XY,c_1,c_2}\left[h_X\left(i,c_1\right),h_Y\left(i,c_2\right)\right] \\
& = \frac{\partial^2}{\partial h_X\left(i,c_1\right)\partial h_Y\left(i,c_2\right)} F_{\mathrm{Copula},c_1,c_2}\left\{\hat{F}_{X,c_1}\left[h_X\left(i,c_1\right)\right],\hat{F}_{Y,c_2}\left[h_Y\left(i,c_2\right)\right]\right\} \\
& = f_{X,c_1}\left[h_X\left(i,c_1\right)\right] f_{Y,c_2}\left[h_Y\left(i,c_2\right)\right] f_{\mathrm{Copula},c_1,c_2}\left\{\hat{F}_{X,c_1}\left[h_X\left(i,c_1\right)\right],\hat{F}_{Y,c_2}\left[h_Y\left(i,c_2\right)\right]\right\}, \\
& = f_{X,c_1}\left[h_X\left(i,c_1\right)\right] f_{Y,c_2}\left[h_Y\left(i,c_2\right)\right] \breve{f}_{\mathrm{Copula},c_1,c_2}\left\{\hat{F}_{X,c_1}\left[h_X\left(i,c_1\right)\right],\breve{F}_{Y,c_1,c_2}\left[h_Y\left(i,c_2\right)\right]\right\}, \forall i,\forall c_1,\forall c_2,
\end{aligned} \tag{31}
$$

where the marginal pdfs of $h_X\left(i,c_1\right)$ and $h_Y\left(i,c_2\right)$ are denoted by $f_{X,c_1}$ and $f_{Y,c_2}$, respectively.

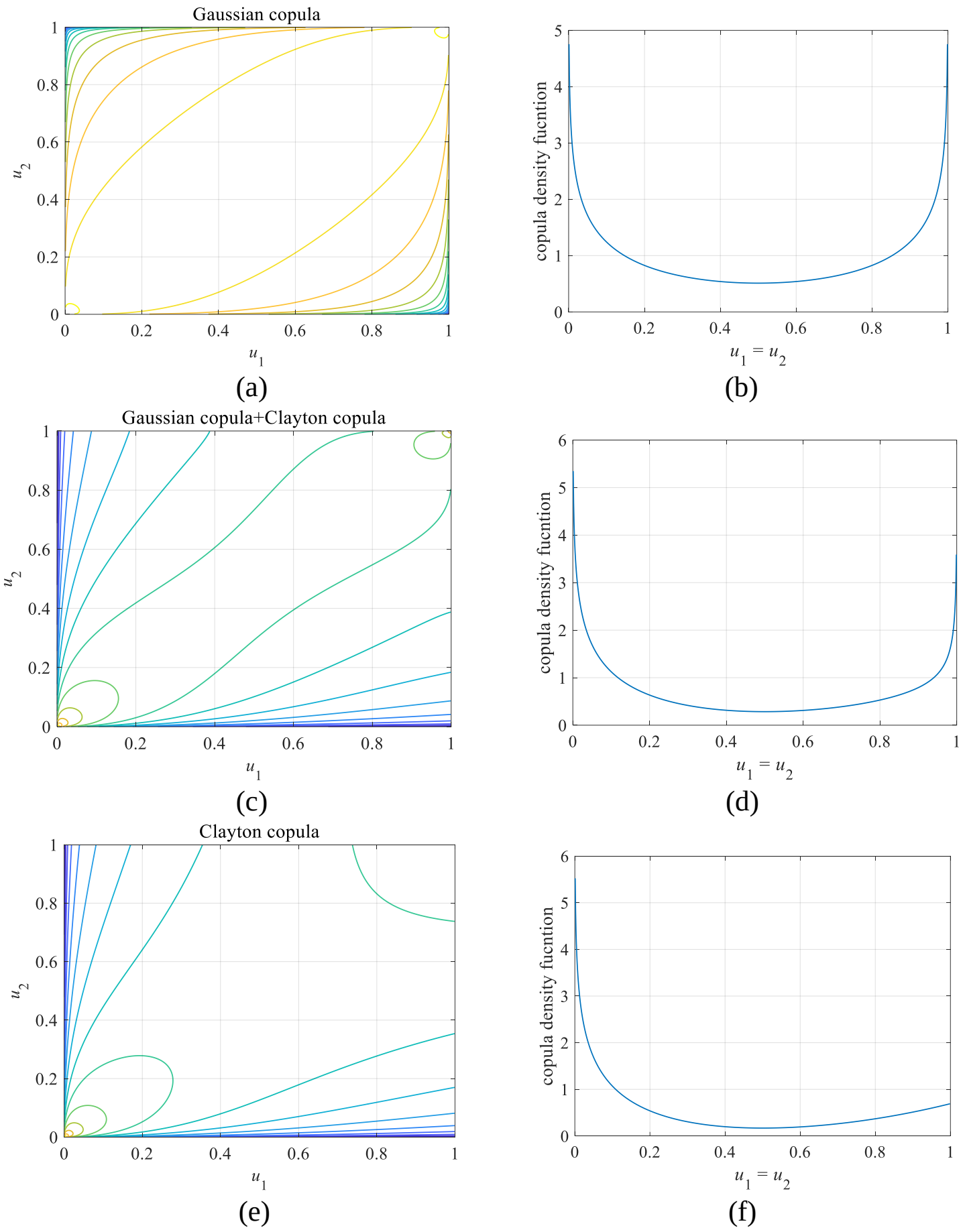

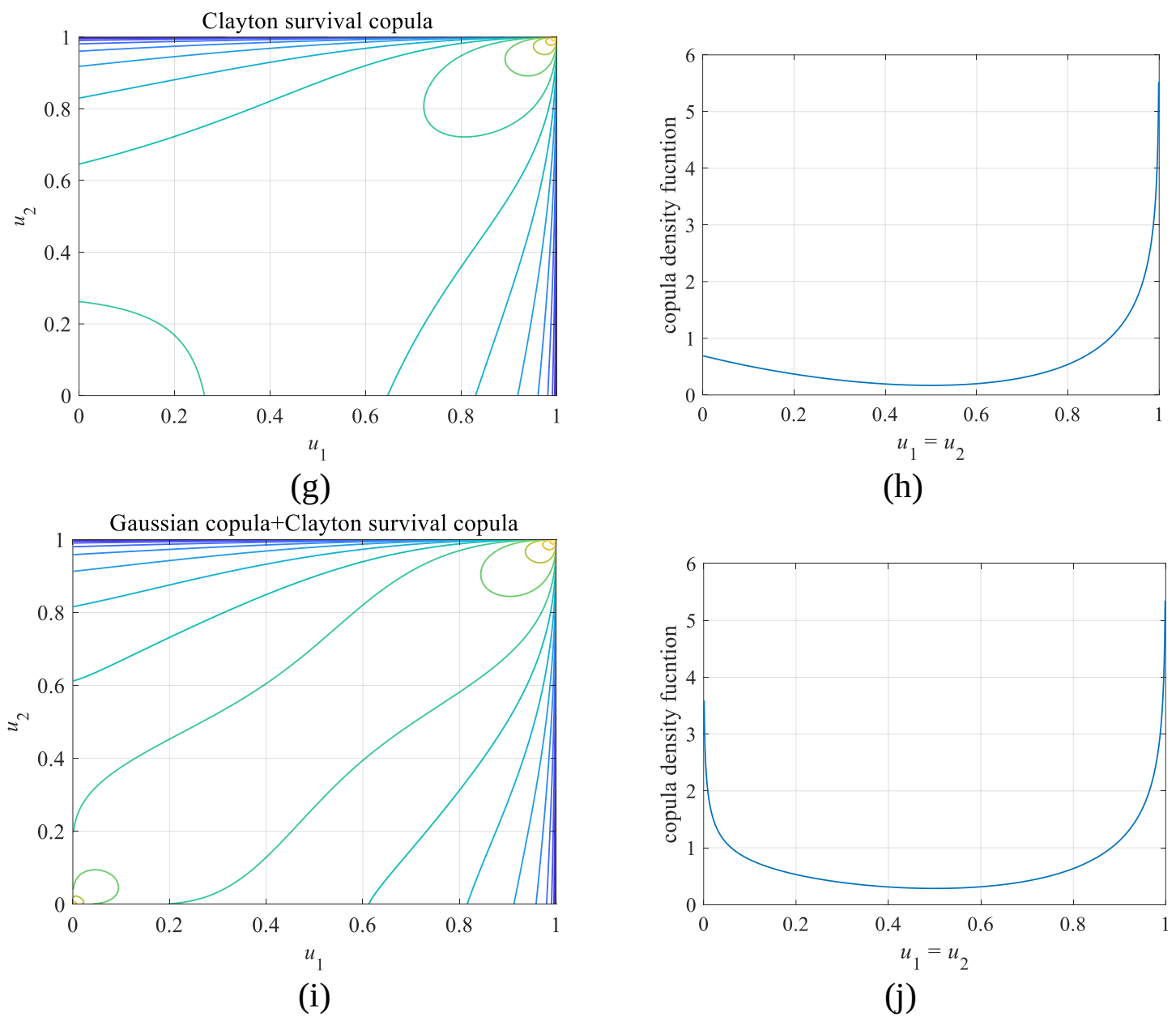

Fig. 3. Contour plots of the copula density functions of the constructed copula mixture and their cross sections on $u_1 = u_2$ , where $\rho_{c_1,c_2} = 0.8$ and $\theta_{c_1,c_2} = 1$. In (a) and (b), the copula mixture equals a Gaussian copula, where $w_{c_1,c_2} = 1$. In (c) and (d), the copula mixture equals a weighted sum of a Gaussian copula and a Clayton copula, where $w_{c_1,c_2} = 0.3$ . In (e) and (f), the copula mixture equals a Clayton copula, where $w_{c_1,c_2} = 0$ . In (g) and (h), the copula mixture equals a Clayton survival copula, where $w_{c_1,c_2} = 0$ . In (i) and (j), the copula mixture equals a weighted sum of a Gaussian copula and a Clayton survival copula, where $w_{c_1,c_2} = 0.3$ .

### *3.4. Parameter estimation*

Next, we show how to estimate the parameters of $\breve{f}_{\text{Copula},c_1,c_2}$ based on the copula mixture constructed in (27) and $\left\{\hat{F}_{X,c_1}\left[h_X\left(i,c_1\right)\right], \breve{F}_{Y,c_1,c_2}\left[h_Y\left(i,c_2\right)\right], \forall i\right\}$ , $\forall c_1$ , $\forall c_2$ . To this end, as done in [53], the samples $\left\{\hat{F}_{X,c_1}\left[h_X\left(i,c_1\right)\right], \breve{F}_{Y,c_1,c_2}\left[h_Y\left(i,c_2\right)\right], \forall i\right\}$ are plugged into the copula density function in (27) to find the Maximum likelihood (ML) estimations of the parameters. For this problem, due to numerical stability and implementation simplicity, EM algorithm is adopted by performing iterative computations, which has been proved to be effective for ML estimation in various problems containing incomplete data [54]. The four steps are presented as follows [54].

Step 1: For channel $c_1$ in image $\mathbf{X}$ and channel $c_2$ in the translated image $\mathbf{Y}'$ , the parameters in the copula mixture are initialized as $\rho^{(0)}_{c_1,c_2}, \theta^{(0)}_{c_1,c_2}, w^{(0)}_{c_1,c_2}$ . We initialize $q = 0$ and set the stopping threshold as $\varepsilon$ . The initial logarithm likelihood function is computed as

$$l\left(\rho^{(0)}_{c_1,c_2}, \theta^{(0)}_{c_1,c_2}, w^{(0)}_{c_1,c_2}\right) = \frac{1}{N_S}\sum_{i=1}^{N_S} \log \breve{f}_{\text{Copula},c_1,c_2}\left[\hat{F}_{X,c_1}\left(h_X\left(i,c_1\right)\right), \breve{F}_{Y,c_1,c_2}\left(h_Y\left(i,c_2\right)\right); \rho^{(0)}_{c_1,c_2}, \theta^{(0)}_{c_1,c_2}, w^{(0)}_{c_1,c_2}\right]. \quad (32)$$

Step 2: Compute

$$\gamma_{i,1}^{(q)} = \frac{w_{c_1,c_2}^{(q)} f_{\text{Gaussian}}\left[\hat{F}_{X,c_1}\left(h_X\left(i,c_1\right)\right), \breve{F}_{Y,c_1,c_2}\left(h_Y\left(i,c_2\right)\right); \rho_{c_1,c_2}^{(q)}\right]}{\breve{f}_{\text{Copula},c_1,c_2}\left[\hat{F}_{X,c_1}\left(h_X\left(i,c_1\right)\right), \breve{F}_{Y,c_1,c_2}\left(h_Y\left(i,c_2\right)\right); \rho_{c_1,c_2}^{(q)}, \theta_{c_1,c_2}^{(q)}, w_{c_1,c_2}^{(q)}\right]}, \gamma_{i,2}^{(q)} = 1 - \gamma_{i,1}^{(q)}, \forall i. \tag{33}$$

Step 3: Compute

$$w_{c_1,c_2}^{(q+1)} = \frac{\sum_{i=1}^{N_S} \gamma_{i,1}^{(q)}}{N_S}, \tag{34}$$

$$\rho_{c_1,c_2}^{(q+1)} = \arg\max_{\rho} \frac{1}{N_S} \sum_{i=1}^{N_S} \gamma_{i,1}^{(q)} \log f_{\text{Gaussian}}\left[\hat{F}_{X,c_1}\left(h_X\left(i,c_1\right)\right), \breve{F}_{Y,c_1,c_2}\left(h_Y\left(i,c_2\right)\right); \rho\right], \tag{35}$$

$$\theta_{c_1,c_2}^{(q+1)} = \begin{cases} \arg\max_{\theta} \frac{1}{N_S} \sum_{i=1}^{N_S} \gamma_{i,2}^{(q)} \log f_{\text{Clayton}}\left[\hat{F}_{X,c_1}\left(h_X\left(i,c_1\right)\right), \breve{F}_{Y,c_1,c_2}\left(h_Y\left(i,c_2\right)\right); \theta\right], & \text{if } \hat{\eta}_{XY,c_1,c_2}^{\text{lower}} > \hat{\eta}_{XY,c_1,c_2}^{\text{upper}}, \\ \arg\max_{\theta} \frac{1}{N_S} \sum_{i=1}^{N_S} \gamma_{i,2}^{(q)} \log f_{\text{S-Clayton}}\left[\hat{F}_{X,c_1}\left(h_X\left(i,c_1\right)\right), \breve{F}_{Y,c_1,c_2}\left(h_Y\left(i,c_2\right)\right); \theta\right], & \text{otherwise.} \end{cases} \tag{36}$$

Note that the solution to (35) and (36) can be found easily by grid search in $(0,1)$ and $\left(0, \theta^{\max}\right)$, respectively, where $\theta^{\max}$ is pre-specified by the system designer.

Step 4: Check the condition of convergence. Compute

$$l\left(\rho_{c_1,c_2}^{(q+1)}, \theta_{c_1,c_2}^{(q+1)}, w_{c_1,c_2}^{(q+1)}\right) = \frac{1}{N_S} \sum_{i=1}^{N_S} \log \breve{f}_{\text{Copula},c_1,c_2}\left[\hat{F}_{X,c_1}\left(h_X\left(i,c_1\right)\right), \breve{F}_{Y,c_1,c_2}\left(h_Y\left(i,c_2\right)\right); \rho_{c_1,c_2}^{(q+1)}, \theta_{c_1,c_2}^{(q+1)}, w_{c_1,c_2}^{(q+1)}\right]. \tag{37}$$

If the following condition holds:

$$\left| l\left(\rho_{c_1,c_2}^{(q+1)}, \theta_{c_1,c_2}^{(q+1)}, w_{c_1,c_2}^{(q+1)}\right) - l\left(\rho_{c_1,c_2}^{(q)}, \theta_{c_1,c_2}^{(q)}, w_{c_1,c_2}^{(q)}\right) \right| < \varepsilon, \tag{38}$$

then stop the iteration. Otherwise, set $q = q+1$ and return to Step 2. When the algorithm converges, the parameters $\rho_{c_1,c_2}^{\text{opt}}, \theta_{c_1,c_2}^{\text{opt}}, w_{c_1,c_2}^{\text{opt}}$ can be obtained.

### *3.5. Test statistic derivation*

In Section 3.2-3.4, we extract features from the translated image and the original pre-event image that are of two different modalities, and we learn the joint distribution of the features from the heterogeneous data by constructing the copula mixture. Note that the translated image is of the same modality as the original post-event image. In this case, the joint distribution of the features extracted from the unchanged area in the two original images can also be characterized by the joint pdf modeled by the constructed copula mixture. On this basis, in Section 3.5, we model the CD problem as a binary hypothesis testing problem and derive its test statistics based on the joint distribution of the features at the superpixel level.

Following similar steps as in Section 3.2, images **X** and **Y** are co-segmented and the resultant co-segmentation map is denoted by $\mathbf{S}_{\text{Test}} \in \mathbb{N}^{M \times N}$, where the number of superpixels is $N_{\text{Test}}$. Based on $\mathbf{S}_{\text{Test}}$, the features extracted from the original images are represented by $\left\{\breve{\mathbf{h}}_X(i), i = 1, ...N_{\text{Test}}\right\}$ and $\left\{\breve{\mathbf{h}}_Y(i), i = 1, ...N_{\text{Test}}\right\}$, where

$$\begin{aligned} \breve{\mathbf{h}}_X(i) &= \left[\breve{h}_X(i,1), ..., \breve{h}_X(i, C_X)\right], \\ \breve{\mathbf{h}}_Y(i) &= \left[\breve{h}_Y(i,1), ..., \breve{h}_Y(i, C_Y)\right]. \end{aligned} \tag{39}$$

With the features extracted from the $i$-th superpixel as shown in (39), for channel $c_1$ in image **X** and channel $c_2$ in image **Y**, we can model the CD problem as a binary hypothesis testing problem as follows based on (31) :

$$\begin{aligned} &\mathrm{H}_0 : \breve{h}_X(i,c_1), \breve{h}_Y(i,c_2) \\ &\sim f_{X,c_1}\left[\breve{h}_X(i,c_1)\right] f_{Y,c_2}\left[\breve{h}_Y(i,c_2)\right] \breve{f}_{\mathrm{Copula},c_1,c_2}\left\{\hat{F}_{X,c_1}\left[\breve{h}_X(i,c_1)\right], \breve{F}_{Y,c_1,c_2}\left[\breve{h}_Y(i,c_2)\right]; \rho^{\mathrm{opt}}_{c_1,c_2}, \theta^{\mathrm{opt}}_{c_1,c_2}, w^{\mathrm{opt}}_{c_1,c_2}\right\}, \\ &\mathrm{H}_1 : \breve{h}_X(i,c_1), \breve{h}_Y(i,c_2) \sim f^1_{X,c_1}\left[\breve{h}_X(i,c_1)\right] f^1_{Y,c_2}\left[\breve{h}_Y(i,c_2)\right], \end{aligned} \tag{40}$$

$c_1 = 1,...,C_X$ , $c_2 = 1,...,C_Y$ , $i = 1,...,N_{\mathrm{Test}}$ , where $\hat{F}_{X,c_1}$ and $\breve{F}_{Y,c_1,c_2}$ are obtained from (17), (18) and (20), $\breve{f}_{\mathrm{Copula},c_1,c_2}$ is the copula density function of the copula mixture constructed in Section 3.3, $\left\{\rho^{\mathrm{opt}}_{c_1,c_2}, \theta^{\mathrm{opt}}_{c_1,c_2}, w^{\mathrm{opt}}_{c_1,c_2}\right\}$ are the parameters estimated in Section 3.4, $\left\{f_{X,c_1}, f_{Y,c_2}\right\}$ denote the marginal pdfs in the unchanged area and $\left\{f^1_{X,c_1}, f^1_{Y,c_2}\right\}$ denote the marginal pdfs in the changed area. In (40), under $\mathrm{H}_0$, no change occurs in the $i$-th superpixel and the features extracted from the images, i.e., $\breve{h}_X(i,c_1)$ and $\breve{h}_Y(i,c_2)$, are statically dependent. In this case, we can denote the joint pdf according to (31) based on the copula mixture constructed in Section 3.3. Under $\mathrm{H}_1$ , changes occur in the $i$-th superpixel, which means the $i$-th superpixel is located in the changed area. In this case, $\breve{h}_X(i,c_1)$ and $\breve{h}_Y(i,c_2)$ are statistically independent, and their joint pdf can be expressed as the product of the marginal pdfs. In this paper, we only consider the case where the type of the objects in the changed area of image **Y** does not depend on the type of the objects in the same location in image **X**, which indicates independence between $\breve{h}_X(i,c_1)$ and $\breve{h}_Y(i,c_2)$ under $\mathrm{H}_1$. Other cases are beyond our consideration in this paper, which will be left for future work.

For the hypothesis testing problem in (40), the test statistics of the optimal log-likelihood ratio (LLR) test are given as

$$\begin{aligned} T_{i,c_1,c_2} &= -\log f_{X,c_1}\left[\breve{h}_X(i,c_1)\right] - \log f_{Y,c_2}\left[\breve{h}_Y(i,c_2)\right] \\ &\quad -\log \breve{f}_{\mathrm{Copula},c_1,c_2}\left\{\hat{F}_{X,c_1}\left[\breve{h}_X(i,c_1)\right], \breve{F}_{Y,c_1,c_2}\left[\breve{h}_Y(i,c_2)\right]; \rho^{\mathrm{opt}}_{c_1,c_2}, \theta^{\mathrm{opt}}_{c_1,c_2}, w^{\mathrm{opt}}_{c_1,c_2}\right\} \\ &\quad +\log f^1_{X,c_1}\left[\breve{h}_X(i,c_1)\right] + \log f^1_{Y,c_2}\left[\breve{h}_Y(i,c_2)\right] \underset{H_0}{\overset{H_1}{\gtrless}} \vartheta, \end{aligned} \tag{41}$$

where $\vartheta$ represents the decision threshold, $\forall i, \forall c_1, \forall c_2$. In this paper, in the absence of prior knowledge regarding the type of the changes, we could make a rational assumption that the marginals of the features in the changed area and in the unchanged area are approximately the same, both in image **X** and image **Y**. This is equivalent to assuming that the objects contained in the changed area and the unchanged area are of similar types. The rationality of the above assumption will be verified by experiments in Section 4. In this case, noting that $f_{X,c_1} \approx f^1_{X,c_1}$ and $f_{Y,c_2} \approx f^1_{Y,c_2}$, the test statistics in (41) can be rewritten as

$$T_{i,c_1,c_2} = -\log \breve{f}_{\mathrm{Copula},c_1,c_2}\left\{\hat{F}_{X,c_1}\left[\breve{h}_X(i,c_1)\right], \breve{F}_{Y,c_1,c_2}\left[\breve{h}_Y(i,c_2)\right]; \rho^{\mathrm{opt}}_{c_1,c_2}, \theta^{\mathrm{opt}}_{c_1,c_2}, w^{\mathrm{opt}}_{c_1,c_2}\right\} \underset{H_0}{\overset{H_1}{\gtrless}} \vartheta. \tag{42}$$

Based on (42), for all the channels in the two images, we can obtain a set of test statistics given as

$$\left\{T_{i,c_1,c_2}, c_1 = 1,...,C_X, c_2 = 1,...,C_Y, i = 1,...,N_{\mathrm{Test}}\right\}. \tag{43}$$

From (42), it can be seen that COMIC identifies the changes by distinguishing between the dependence structure captured by the constructed copula mixture and the independence structure between the features extracted from the original heterogeneous images.

*3.6. Difference map and BCM calculation*

Based on the test statistics in (43), the final difference map $\mathrm{DI}=\{\mathrm{DI}_i,\forall i\}$ can be obtained in the following way:

$$\mathrm{DI}_i=\max\{T_{i,c_1,c_2},\forall c_1,\forall c_2\},\forall i. \tag{44}$$

The fusion rule in (44) effectively reduces the false negatives of CD by highlighting all the changes identified by any two channels of the heterogeneous images in the final difference map. After obtaining the final difference map DI, we show how to generate the BCM by means of two-stage K-means clustering, leveraging the information provided by the original images and the final difference map.

Let us construct a representative vector for the $i$-th superpixel:

$$\mathbf{r}_i=\left[\breve{h}_X(i,1),...,\breve{h}_X(i,C_X),\breve{h}_Y(i,1),...,\breve{h}_Y(i,C_Y),\alpha\tilde{\mathrm{D}}\mathrm{I}_i\right],\forall i, \tag{45}$$

where

$$\tilde{\mathrm{D}}\mathrm{I}_i=\mathrm{DI}_i-\kappa(\mathrm{DI}),\forall i, \tag{46}$$

$\kappa(\bullet)$ is the mean function, and $\alpha$ is the parameter that controls the weight of the difference map during the process of clustering. In the first stage, $\{\mathbf{r}_i,i=1,...,N_{\mathrm{Test}}\}$ are clustered into three clusters $\{\mathcal{A}_1,\mathcal{A}_2,\mathcal{A}_3\}$ by K-means clustering, including one cluster $\mathcal{A}_1$ containing $Q_1$ significantly unchanged superpixels, one cluster $\mathcal{A}_2$ containing $Q_2$ significantly changed superpixels and one cluster $\mathcal{A}_3$ containing $Q_3$ superpixels that are neither significantly changed nor significantly unchanged. Among the three clusters, cluster $\mathcal{A}_2$ can be identified by finding the cluster in which the superpixels have larger values in the difference map. Equivalently, $\mathcal{A}_2$ is given as

$$\mathcal{A}_2=\underset{\mathcal{A}\subset\{\mathcal{A}_1,\mathcal{A}_2,\mathcal{A}_3\}}{\arg\max}\frac{1}{|\mathcal{A}|}\sum_{i\in\mathcal{A}}\mathrm{DI}_i. \tag{47}$$

In the second stage, $\{\mathbf{r}_i,i=1,...,N_{\mathrm{Test}}\}$ are clustered into two clusters $\{\mathcal{B}_1,\mathcal{B}_2\}$ by means of K-means clustering, where $\mathcal{B}_1$ contains the superpixels in the unchanged area and $\mathcal{B}_2$ contains the superpixels in the changed area. The cluster $\mathcal{B}_2$ can be identified by finding the cluster which has a larger overlap with $\mathcal{A}_2$, which means $\mathcal{B}_2$ contains more significantly changed superpixels. Equivalently, $\mathcal{B}_2$ is given as

$$\mathcal{B}_2=\underset{\mathcal{B}\subset\{\mathcal{B}_1,\mathcal{B}_2\}}{\arg\max}|\mathcal{A}_2\cap\mathcal{B}|. \tag{48}$$

Finally, the BCM $\mathbf{B}\sim\mathbb{N}^{M\times N}$ is generated as

$$\begin{aligned}&b(m,n)=1,\ \text{if } s_{\mathrm{Test}}(m,n)=i,i\in\mathcal{B}_2,\\&b(m,n)=0,\ \text{otherwise},\end{aligned} \tag{49}$$

where the $(m,n)$-th element in $\mathbf{B}$ is denoted by $b(m,n)$ and the $(m,n)$-th element in $\mathbf{S}_{\mathrm{Test}}$ is denoted by $s_{\mathrm{Test}}(m,n)$.

## 4. Experimental results

To validate the superiority of COMIC, we run experiments on real RS datasets which have been widely adopted to demonstrate the CD performance of different methods in the recent literature [19, 62, 72]. Dataset 1 contains two images including a pre-event image of size $300\times412\times1$ and a post-event image of size $300\times 412\times 3$. They were acquired by Landsat-5 in Sept. 1995 and by Google Earth in Jul. 1996, respectively. The monitored event is a lake expansion in Sardinia, Italy. Dataset 2 contains two images, where the pre-event image of size $990\times 554\times 3$ was acquired by Spot in 1999 and the post-event image of size $990\times 554\times 1$ was acquired by NDVI in 2000. The monitored event is a flooding event in Gloucester, England. Dataset 3 contains two multispectral images taken in Bastrop County, Texas, USA [72]. The pre-event image of size $1534\times 808\times 7$ was acquired by Landsat 5 in September 2011 before a forest fire. The post-event image of size $1534\times 808\times 10$ was acquired by Earth Observing-1 Advanced Land Imager in October 2011. In each dataset, the two images are registered before further processing.

We adopt three quantitative performance measures to evaluate the CD performance explicitly: Kappa coefficient (KC), F-measure (Fm) and overall accuracy (ACC), which are expressed as

$$\begin{aligned} \mathrm{KC} &= \frac{2\left(TP\times TN-FN\times FP\right)}{\left(TP+FP\right)\times\left(FP+TN\right)+\left(TP+FN\right)\times\left(FN+TN\right)}, \\ \mathrm{Fm} &= \frac{2TP}{2TP+FP+FN}, \\ \mathrm{ACC} &= \frac{TP+TN}{TP+TN+FP+FN}. \end{aligned} \tag{50}$$

In (50), True Positive, True Negative, False Positive and False Negative are denoted by *TP*, *TN*, *FP*, and *FN*, respectively. It is worth mentioning that larger values of KC, Fm and ACC indicate enhanced CD performance. During the process of image translation, unpaired training samples are collected by a $64\times 64$ sliding window with step size $\lambda=8$ for Dataset 1, by a $128\times128$ sliding window with step size $\lambda=16$ for Dataset 2, and by a $64\times 64$ sliding window with step size $\lambda=32$ for Dataset 3. The parameters are initialized as $\rho_{c_1,c_2}^{(0)}=0.5,\theta_{c_1,c_2}^{(0)}=0.5$ and $w_{c_1,c_2}^{(0)}=0.5$ when EM algorithm is implemented. The stopping criterion of EM algorithm is set as $\varepsilon=0.01$.

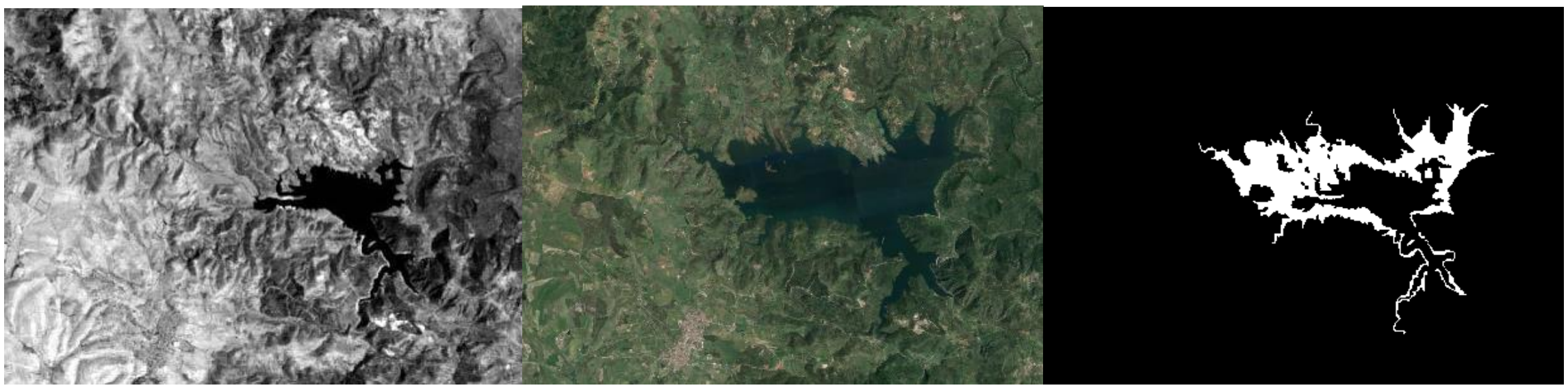

(a) (b) (c)

Fig. 4. Dataset 1. (a) The pre-event image; (b) The post-event image; (c) Ground truth.

### *4.1. Experiments on Dataset 1*

The pre-event image **X**, the post-event image **Y**, and the ground truth are shown in Fig. 4(a)-(c). The original pre-event image and the translated image are co-segmented into $N_S=1000$ superpixels. The original pre-event image and the post-event image are co-segmented into $N_{\text{Test}}=2000$ superpixels. In K-means clustering, we fix $\alpha=5$.

First, we compare the CD performances of COMIC and ITS [19]. For a fair comparison, the same translated image is utilized by both methods. When ITS is implemented, the difference map is obtained by calculating the pixelwise distances between the translated image $\mathbf{Y}'$ and the original image $\mathbf{Y}$ via direct subtraction, both of which are of the same modality. Note that the two images contain three channels. In this case, the pixelwise distances in the difference map are calculated as the mean values of the absolute values of the differences in three channels, as done in [19]. When COMIC is implemented, the difference map is generated according to the procedure introduced in Section 3 in this paper. After the difference map is obtained, for a fair comparison, BCM is generated in the same way for both COMIC and ITS as introduced in Section 3.6 at the superpixel level and the optimal number of clusters in the second stage of K-means clustering is adopted for ITS. It is worth noting that how to generate the BCM is not the main focus of this work. Other alternative techniques could be adopted for BCM generation, which does not impact the main contributions in this paper. Actually, the main difference between COMIC and ITS lies in the fact that the former calculates the difference map based on copula construction and statistical modelling, while the latter performs direct subtraction. The experiments in this paper eliminate the influence of BCM generation on CD performance and enable us to focus on the influence of different approaches of difference map calculation adopted by different CD methods. Based on the fact that COMIC and ITS use the same data generated by CycleGANs and utilize the same technique for BCM generation, comparing the performances of COMIC and ITS results in an ablation study which investigates the impact of adopting copula mixtures. Fig. 5(a) and (b) present the difference maps obtained by COMIC and ITS. We can easily observe that the changed area in Fig. 5(a) is more discriminative compared with that in Fig. 5(b), and the quality of the difference map of the proposed method is much better than that of ITS. Correspondingly, in Fig. 5(c) and (d), we provide the histograms of the values in the normalized difference map of the two methods. It can be seen that the values in the difference map in the changed area and the unchanged area are more separable when COMIC is implemented, which is consistent with our observations in Fig. 5(a) and (b). In this case, COMIC makes it easy to distinguish between the changed area and the unchanged area. The rationale behind this phenomenon is that ITS is sensitive to image translation errors in the unchanged area while COMIC has a better tolerance to them. For example, in the difference map of ITS, large values are assigned to the pixels in the black box in Fig. 5(b) due to image translation errors existing in the unchanged area. In this case, the difference map of ITS erroneously indicates that changes are very likely to occur in this area. In contrast, small values are assigned to the pixels in the black box in the difference map of COMIC in Fig. 5(a), which contributes to correctly identifying this area as the unchanged area. The BCMs of the two methods are shown in Fig. 5(e) and (f), which demonstrates that COMIC attains superior CD performance with less false positives and false negatives compared with ITS.

To further validate the superiority of COMIC, we compare the CD performances of COMIC and the state-of-the-art CD methods[4], which include iterative robust graph and Markovian co-segmentation method with distance criterion (IRG-McS.dist) [20], iterative robust graph and Markovian co-segmentation method with similarity criterion (IRG-McS.sim) [20], conditional copula-based CD technique (C3D) [22], anomaly feature learning based deep sparse residual model (AFL-DSR) [18], Markov model for multimodal CD method (M3CD) [63], ITS [19], deep image translation with affinity-based prior (ITA) [59], deep convolutional coupling network (SCCN) [17], conditional adversarial network (CAN) [60], non-local patch similarity-based graph (NPSG) [15], code-aligned autoencoders (CAA) [61], improved non-local patch-based graph (INLPG) [21], and topology-coupling-based heterogeneous RS image change detection network (TSCNet) [62]. The experimental results of IRG-McS.dist, IRG-McS.sim, AFL-DSR, M3CD and NPSG are quoted from [20] and the experimental results of ITA, SCCN, CAN, CAA, INLPG and TSCNet are quoted from [62]. For C3D, Ali-Mikhail-Haq copula is utilized on account of its performance gain stated in [22] and its tractable form for implementation. The Kullback-Leibler divergence is calculated by using Predictive Maintenance Toolbox of

[4] As done in [20], the results of the state-of-the-art methods are directly quoted from published papers. Based on the fact that different methods are tested on different datasets in the literature, in this paper, different state-of-the-art methods are adopted for performance comparison on different datasets.

MATLAB. The quantitive performance measures of different methods are presented in Table 1. By comparing the quantitive performance measures of COMIC and the state-of-the-art methods in Table 1, it can be verified that COMIC outperforms other methods in terms of the CD performance. It is worth emphasizing that, even though COMIC only slightly outperforms IRG-McS.dist and IRG-McS.sim in terms of the ACC values as shown in Table 1, we can observe that COMIC attains higher values of KC and Fm, which indicates that more false positives and false negatives are generated by IRG-McS.dist and IRG-McS.sim compared with COMIC. From this perspective, the superiority of COMIC is quite evident. The reasons behind the superiority of the proposed method can be stated as follows. First, the proposed method fully exploits the advantages of CycleGANs on data mining and produces sufficient data where the joint distribution and the dependence structure between the features of the heterogeneous RS images can be learnt. Second, COMIC utilizes copula theory to decouple the marginal distributions and the dependence structure between the features of the heterogeneous RS images, which separates the complementary information and the common information contained in the multi-temporal images. This contributes to more efficient fusion of the information contained in the heterogeneous RS data. Third, COMIC constructs copula mixtures to characterize various dependence structures between the features of the heterogeneous RS data in a more flexible and accurate way, which further enhances the CD performance.

The sensitivity of COMIC to the number of superpixels $N_S$ is well worth discussing in order to verify its robustness. To this end, Fig. 6 plots the quantitative performance measures under different values of $N_S$. It can be seen that the CD performance is relatively constant, which verifies that the proposed method is quite robust to the changes in the number of superpixels.

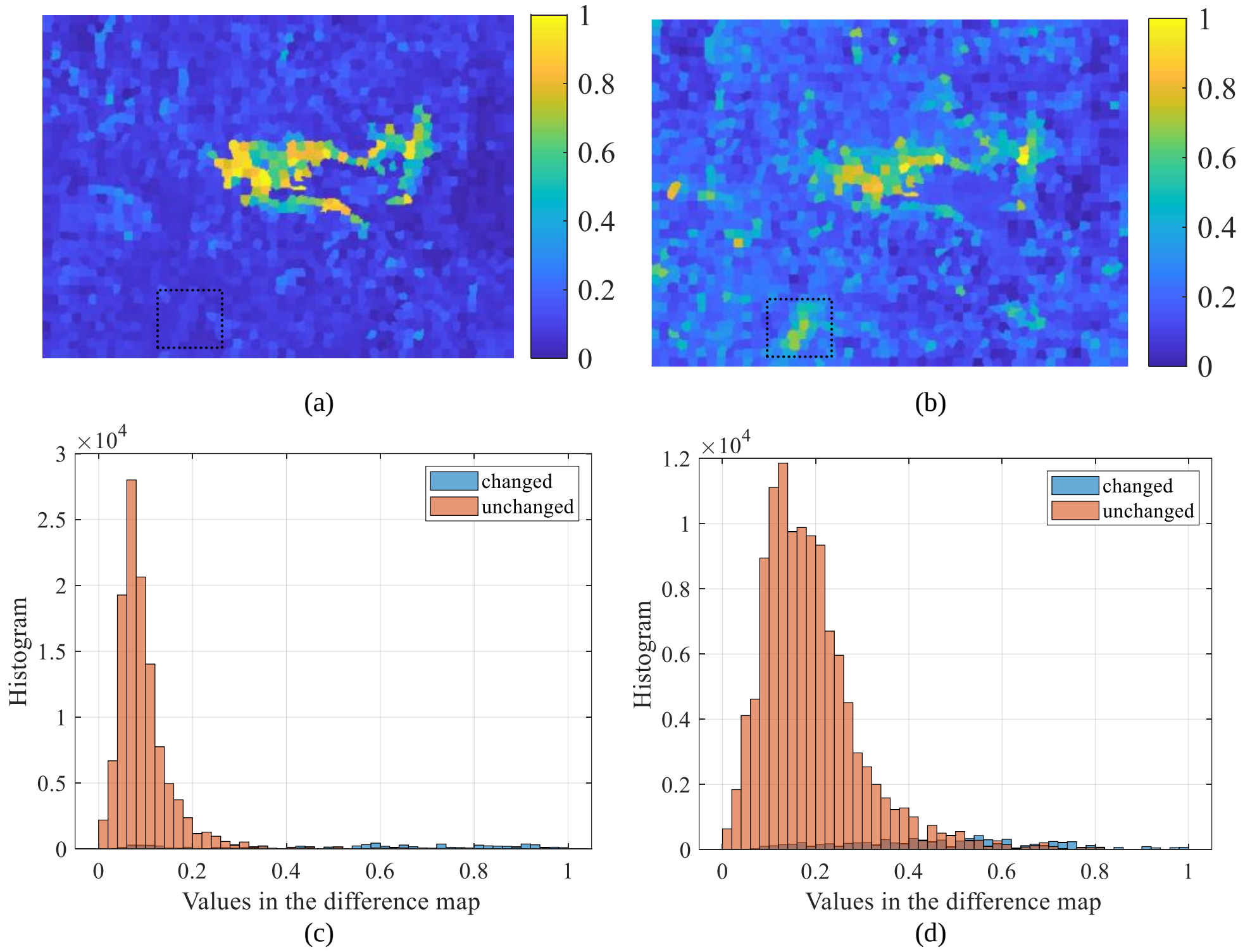

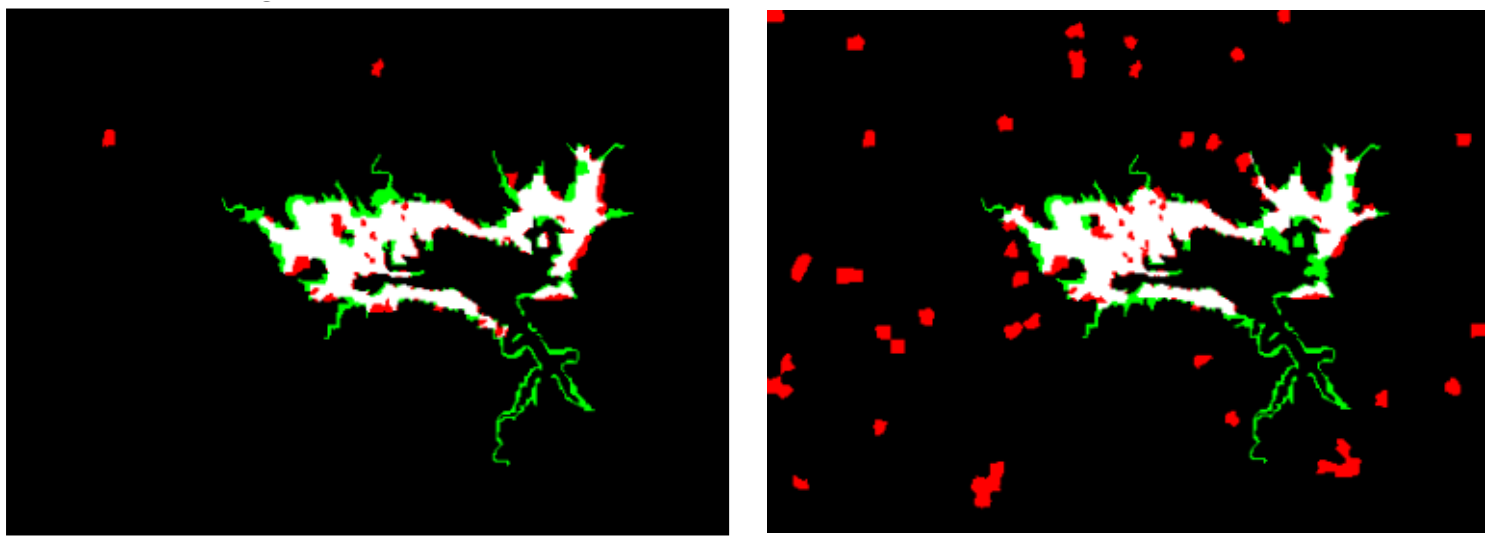

(e) (f)

TP FP TN FN

Fig. 5. The CD performances of COMIC and the state-of-the-art methods on Dataset 1. (a) The difference map of COMIC; (b) The difference map of ITS [19]; (c) The histogram of the values in the difference map of COMIC; (d) The histogram of the values in the difference map of ITS; (e) The BCM of COMIC; (f) The BCM of ITS.

Table 1. Quantitative performance measures on Dataset 1.

| Methods | KC | Fm | ACC |
|---|---|---|---|
| IRG-McS.dist | 0.739 | 0.754 | 0.971 |
| IRG-McS.sim | 0.733 | 0.749 | 0.971 |
| C3D | - | - | 0.925 |
| AFL-DSR | - | - | 0.929 |
| M3CD | 0.669 | 0.689 | 0.963 |
| ITS | 0.635 | 0.660 | 0.954 |
| ITA | 0.362 | 0.412 | 0.902 |
| SCCN | 0.412 | 0.462 | 0.892 |
| CAN | 0.401 | 0.439 | 0.929 |
| NPSG | 0.559 | 0.587 | 0.947 |
| CAA | 0.631 | 0.658 | 0.949 |
| INLPG | 0.506 | 0.546 | 0.918 |
| TSCNet | 0.654 | 0.676 | 0.955 |
| **The proposed** | **0.761** | **0.774** | **0.974** |

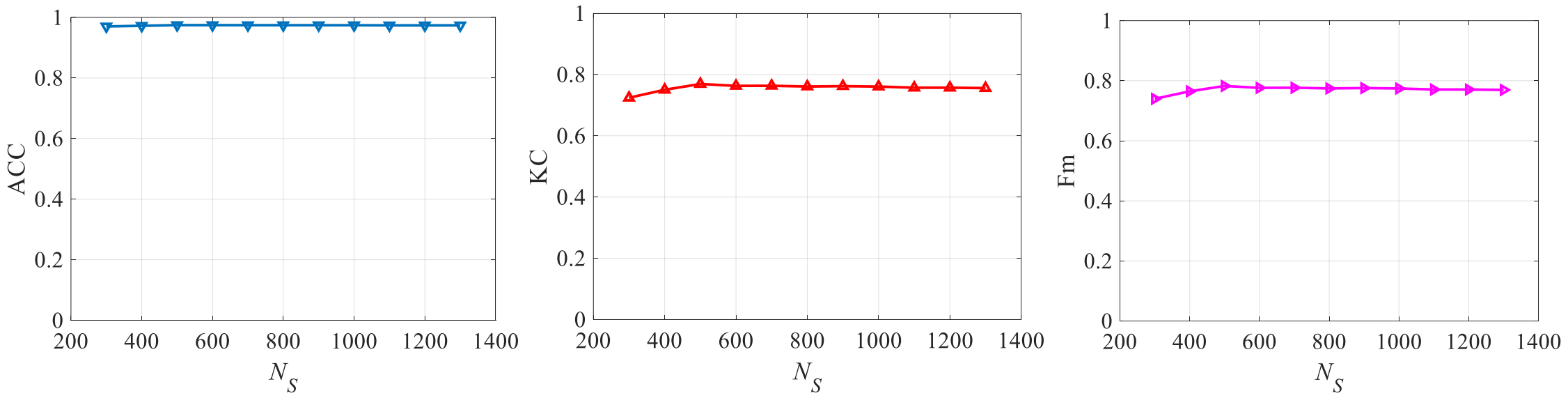

(a) (b) (c)

Fig. 6. The quantitative performance measures on Dataset 1 under different values of $N_S$. (a) ACC values; (b) KC values; (c) Fm values.

### *4.2. Experiments on Dataset 2*

The pre-event image, the post-event image, and the ground truth in Dataset 2 are shown as Fig. 7(a)-(c), respectively. The original pre-event image and the translated image are co-segmented into $N_S = 4000$ superpixels, and the original images are co-segmented into $N_{\text{Test}} = 4000$ superpixels. In K-means clustering, we fix $\alpha = 5$. The CD performances of COMIC and ITS on Dataset 2 are compared in Fig. 8. The difference maps obtained by COMIC and ITS are shown in Fig. 8(a) and (b), and the histograms of the values in the normalized difference maps of the two methods are shown in Fig. 8(c) and (d). From Fig. 8(a)-(d), it can be seen that, by performing COMIC, a clearer difference map can be obtained, where the changed area and the unchanged area are more distinguishable. Accordingly, the BCMs of COMIC and ITS are presented in Fig. 8(e) and (f), which demonstrates the superior CD performance of COMIC over ITS.

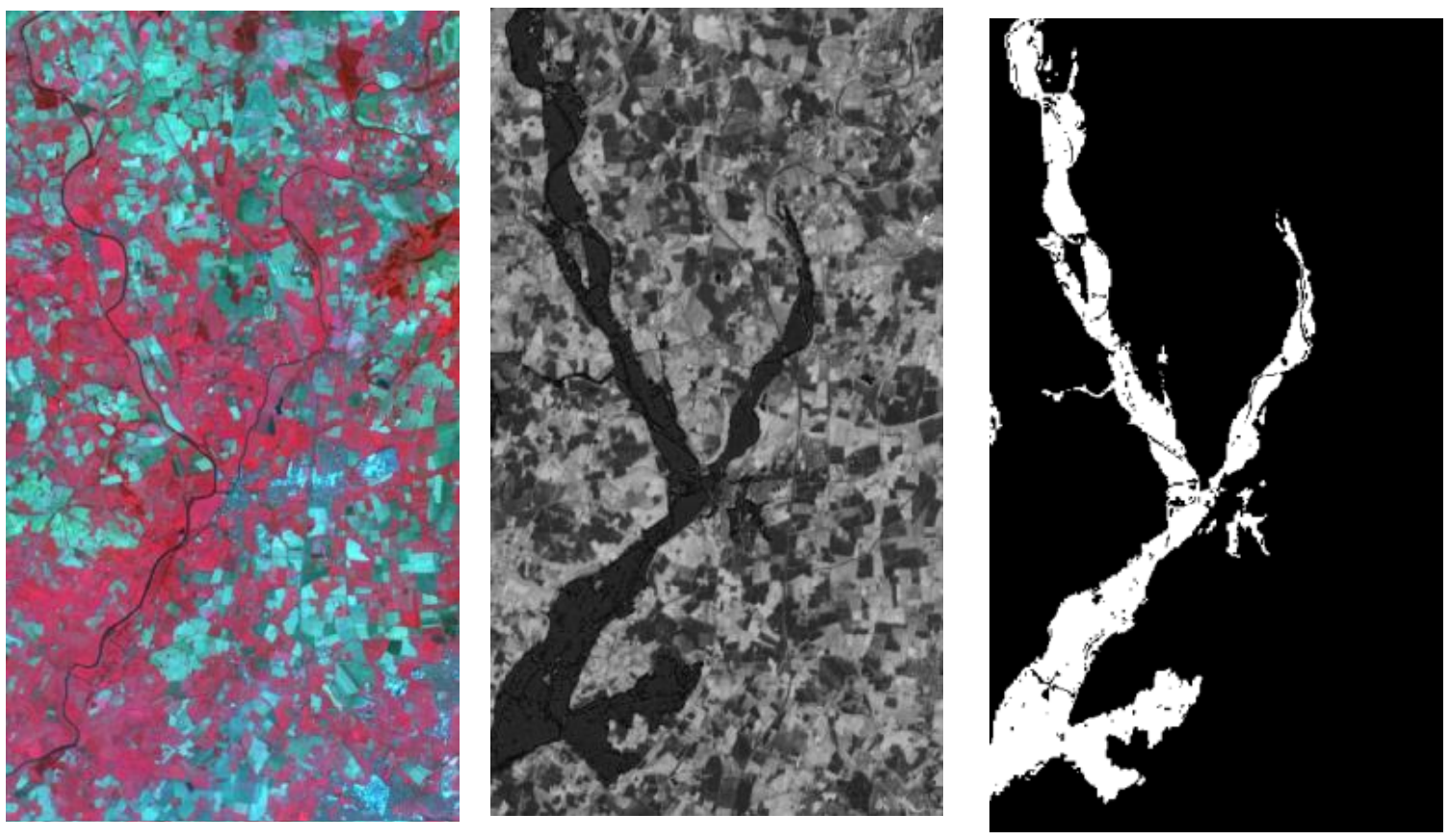

(a) (b) (c)

Fig. 7. Dataset 2. (a) The pre-event image; (b) The post-event image; (c) Ground truth.

Additionally, in Table 2, we compare the CD performances of COMIC and the state-of-the-art methods, which include IRG-McS.dist [20], IRG-McS.sim [20], C3D [22], AFL-DSR [18], ITS [19], M3CD [63], NPSG [15], multidimensional evidential reasoning based method (MDER) [64], adaptive local structure consistency-based method (ALSC) [65], patch similarity graph matrix-based method (PSGM) [66], and sparse-constrained adaptive structure consistency-based image regression method for heterogeneous CD (SCASC) [67]. The experimental results of IRG-McS.dist, IRG-McS.sim, AFL-DSR, M3CD, NPSG and MDER are quoted from [20]. The experimental results of ALSC, PSGM and SCASC are quoted from [67]. C3D is implemented as introduced in Section 4.1. From these results on Dataset 2, the superiority of COMIC in terms of the CD performance is further validated. We would like to note that, among the proposed method and the state-of-the-art methods, the computational complexities of COMIC, ITS and AFL-DSR are higher than others, which is due to the fact that their implementations involve training deep models. However, we should also note that training such models can be performed as off-line tasks, which means computational complexity is not a dominant factor in operational scenarios.

Table 2. Quantitative performance measures on Dataset 2.

| Methods | KC | Fm | ACC |
|---|---|---|---|
| IRG-McS.dist | 0.714 | 0.749 | 0.939 |
| IRG-McS.sim | 0.735 | 0.768 | 0.942 |
| C3D | - | - | 0.899 |
| AFL-DSR | - | - | 0.836 |
| ITS | 0.687 | 0.723 | 0.937 |
| M3CD | 0.588 | 0.636 | 0.915 |
| NPSG | 0.608 | 0.663 | 0.902 |
| MDER | - | - | 0.818 |
| ALSC | 0.641 | 0.693 | 0.907 |
| PSGM | 0.675 | 0.719 | 0.922 |
| SCASC | 0.771 | 0.800 | 0.949 |
| **The proposed** | **0.807** | **0.830** | **0.960** |

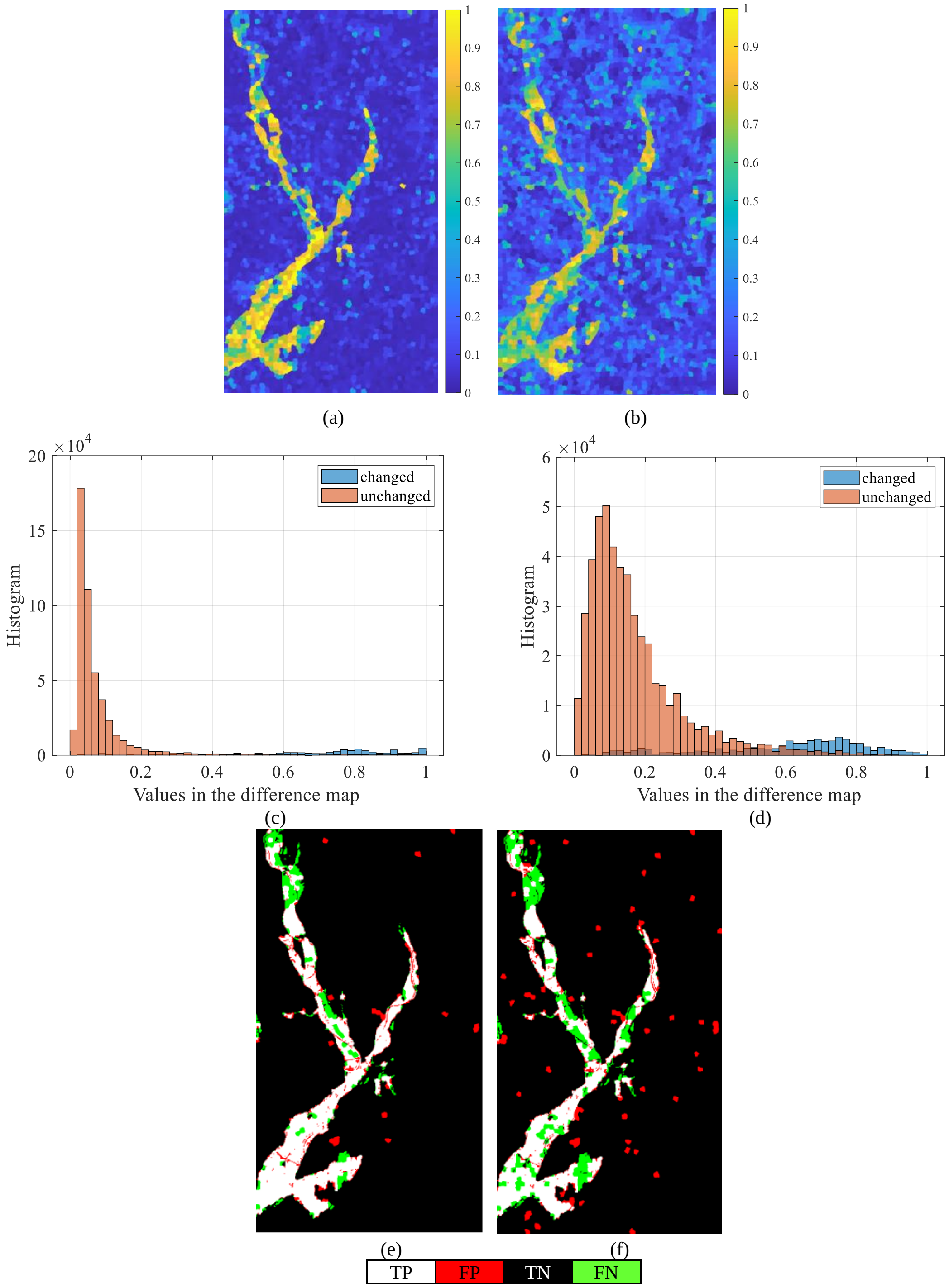

Fig. 8. The CD performances of COMIC and the state-of-the-art methods on Dataset 2. (a) The difference map of COMIC; (b) The difference map of ITS [19]; (c) The histogram of the values in the difference map of COMIC; (d) The histogram of the values in the difference map of ITS; (e) The BCM of COMIC; (f) The BCM of ITS.

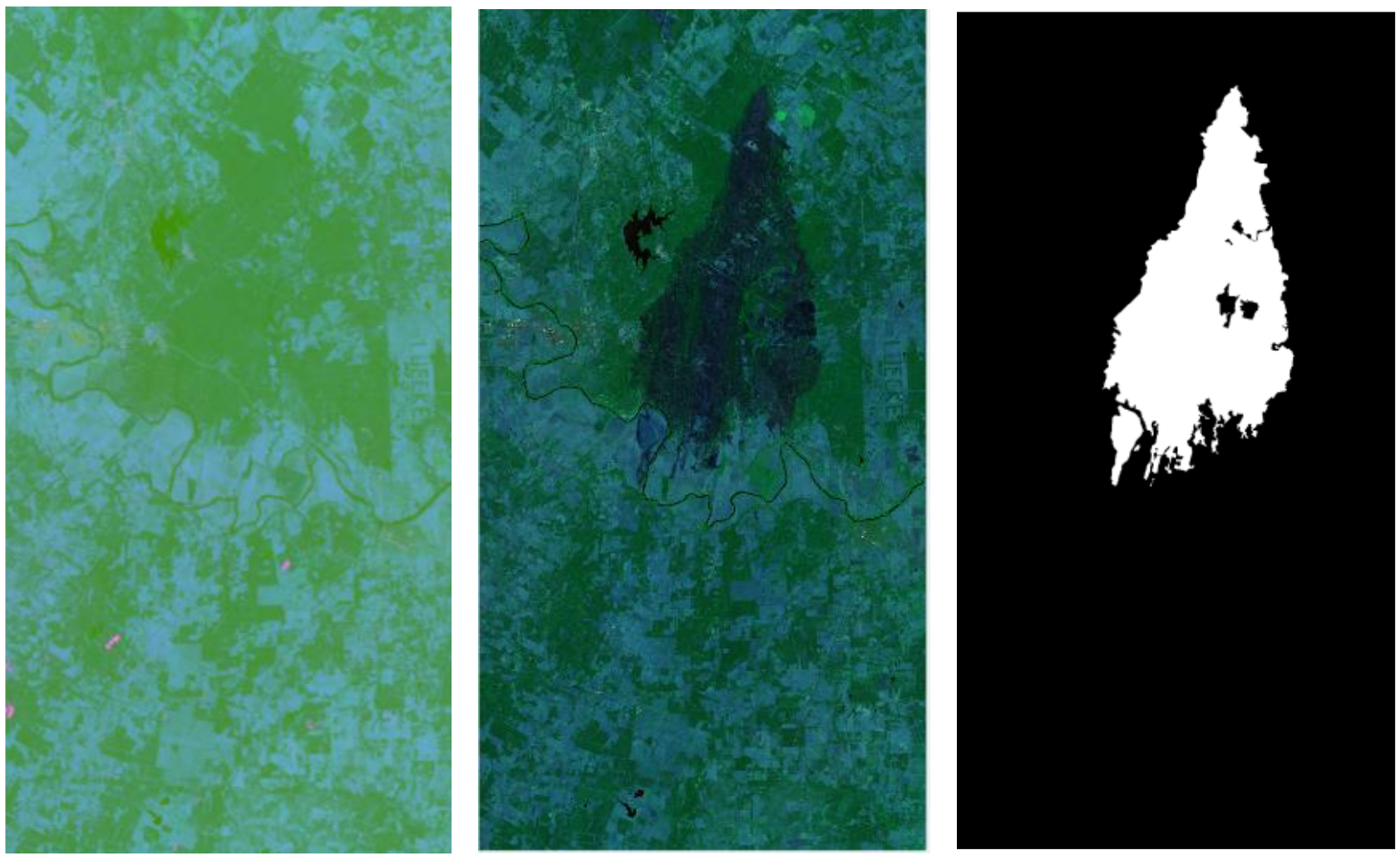

(a) (b) (c)

Fig. 9. Dataset 3. (a) The pre-event image; (b) The post-event image; (c) Ground truth.

### *4.3. Experiments on Dataset 3*

In Fig. 9(a)-(c), the pseudo-color pre-event image, the pseudo-color post-event image, and the ground truth in Dataset 3 are presented. The original pre-event image and the translated image are co-segmented into $N_S = 900$ superpixels, and the two original images are co-segmented into $N_{\text{Test}} = 1800$ superpixels. In K-means clustering, we set $\alpha = 0.03$. Considering that the original pre-event image and the post-event image have a large number of bands, to reduce the complexity of image translation, spectral dimensionality reduction is performed as a preprocessing step by computing a specified number of principal components from the original multispectral images. For both the pre-event image and the post-event image, two principal component bands are adopted. In Fig. 10, we compare the CD performances of COMIC and ITS achieved on Dataset 3, where the difference maps and the histograms of the values in the normalized difference maps of the two methods are presented. It can be observed from Fig. 10 that COMIC obtains a clearer difference map than ITS, and also attains superior CD performance.

The quantitative performance measures of different methods are shown in Table 3, where the state-of-the-art methods include ITS [19], ITA [59], SCCN [17], CAN [60], NPSG [15], CAA [61], INLPG [21], and TSCNet [62]. The experimental results of ITA, SCCN, CAN, NPSG, CAA, INLPG, and TSCNet are quoted from [62]. From Table 3, it can be clearly observed that higher values of KC, Fm and ACC are attained by COMIC than the state-of-the-art methods, which demonstrates the superiority of COMIC. Although COMIC only slightly outperforms CAA and TSCNet based on the results presented in Table 3, it is worth noting that the proposed method significantly improves the interpretability of the deep learning methods including CAA and TSCNet. In that case, the reasoning behind the results of the proposed method is more easily comprehensible and COMIC is much trustier than others, which validates the value of our work.

Table 3. Quantitative performance measures on Dataset 3.

| Methods | KC | Fm | ACC |
|---|---|---|---|
| ITS | 0.847 | 0.864 | 0.971 |
| ITA | 0.701 | 0.730 | 0.948 |
| SCCN | 0.694 | 0.734 | 0.928 |
| CAN | 0.576 | 0.615 | 0.929 |
| NPSG | 0.612 | 0.666 | 0.895 |
| CAA | 0.871 | 0.885 | 0.975 |
| INLPG | 0.582 | 0.618 | 0.932 |
| TSCNet | 0.873 | 0.879 | 0.976 |
| **The proposed** | **0.878** | **0.890** | **0.977** |

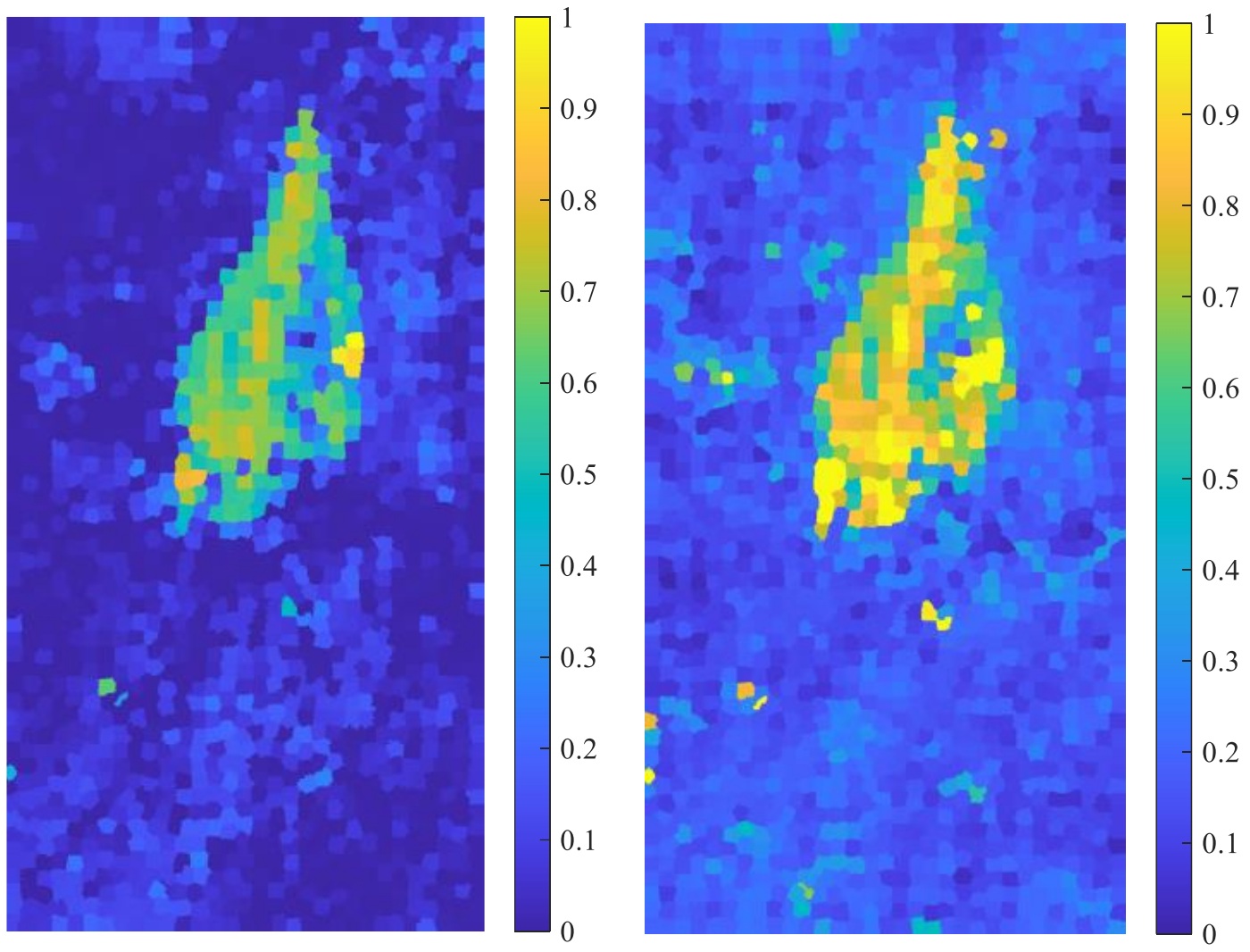

(a) (b)

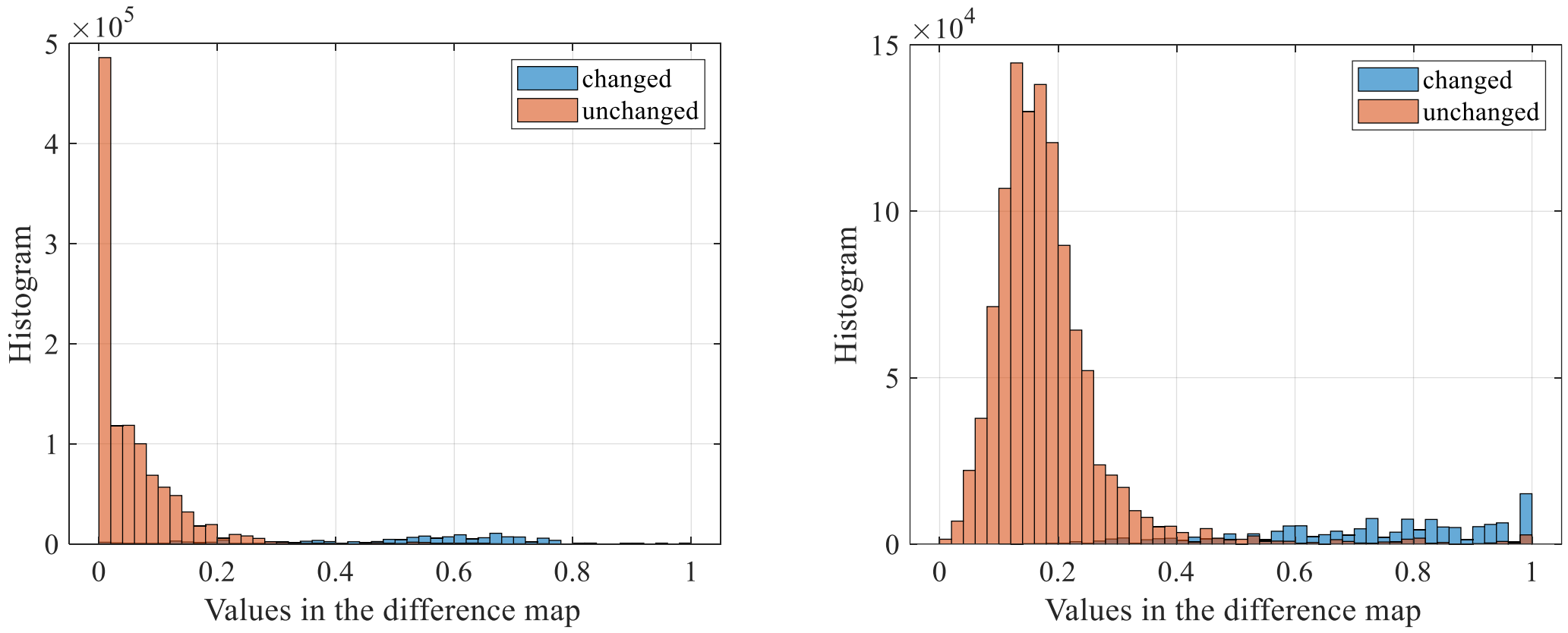

(c) (d)

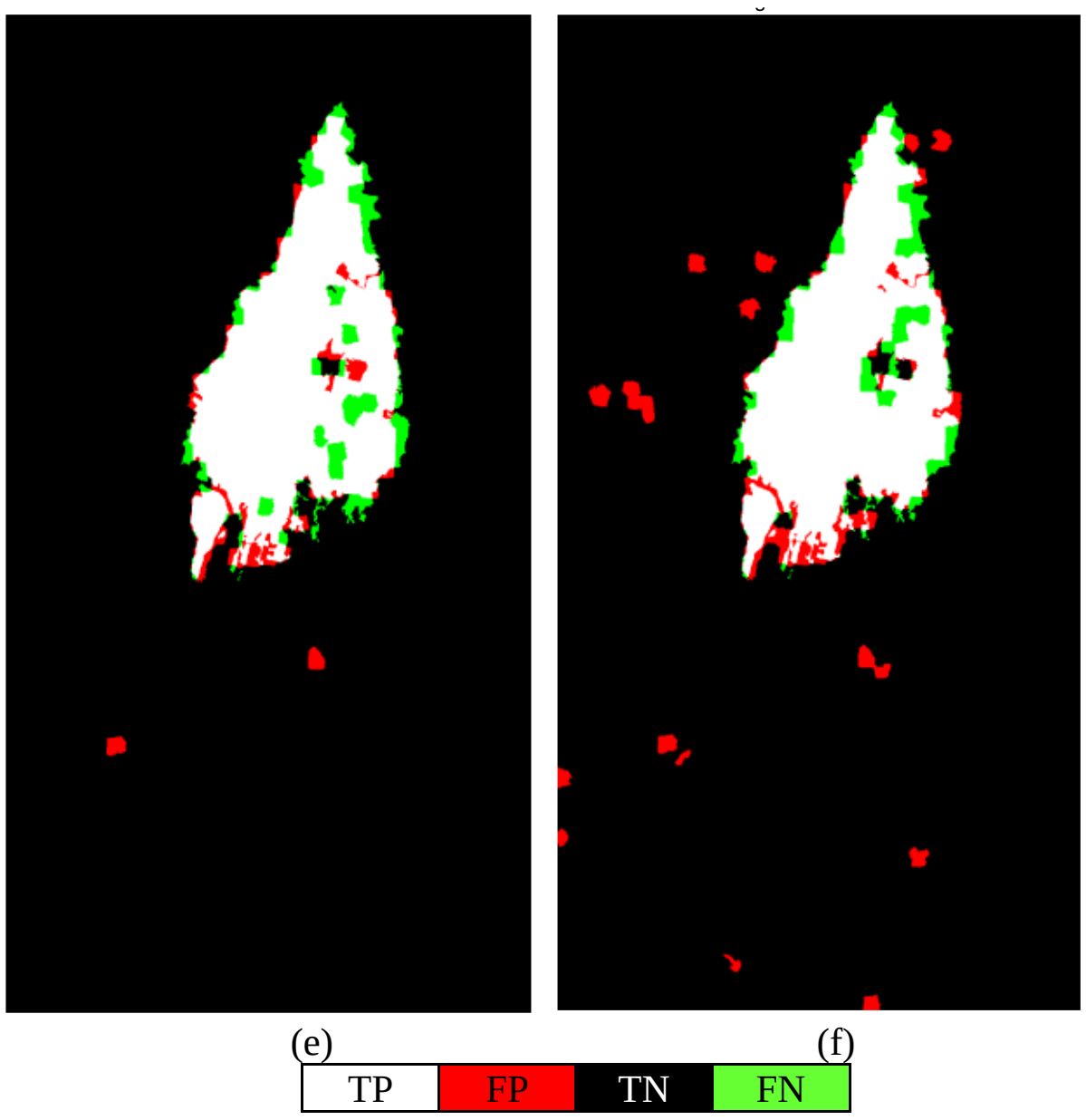

Fig. 10. The CD performances of COMIC and the state-of-the-art methods on Dataset 3. (a) The difference map of COMIC; (b) The difference map of ITS [19]; (c) The histogram of the values in the difference map of COMIC; (d) The histogram of the values in the difference map of ITS; (e) The BCM of COMIC; (f) The BCM of ITS.

## 5. Conclusion

In this paper, the CD problem with heterogeneous RS images was investigated. Considering that ITS proposed in the literature is sensitive to image translation errors in the unchanged area, a new unsupervised CD method named COMIC was proposed to overcome the drawback of ITS by combining the advantages of copula mixtures and CycleGANs. As the first step of COMIC, the pre-event image was translated from its original modality to the post-event image modality. After that, the copula mixture was constructed to model the joint distribution of the features extracted from the heterogeneous images, including the original pre-event image and the translated image, according to quantitive measures of the dependence structure, and the parameters in the copula mixture were estimated iteratively by EM algorithm. Then, the CD problem was modeled as a binary hypothesis testing problem, and we derived the test statistics based on the constructed copula mixture. Finally, the difference map was obtained based on the test statistics and the BCM was generated by two-stage K-means clustering. Experiments were conducted on real RS datasets, which demonstrated the superiority of COMIC over the state-of-the-art methods. Specifically, when compared to ITS, COMIC achieves an increase in ACC ranging from 0.6% to 2%, an enhancement in KC from 0.031 to 0.126, and an improvement in Fm from 0.026 to 0.114 on different datasets that we considered. This level of improvement is noteworthy given the high performance achieved by existing methods. In fact, even marginal gains in CD performance can have substantial impacts on the reliability and effectiveness of CD, making our findings significant for advancing this field.

It is worth noting that in this work, the CD task is implemented as a fusion task at the superpixel level, which fuses information contained in the multi-temporal RS images by adopting superpixels as the basic processing units. In the future, we would extend the present work by extracting more complicated features at different levels from the heterogeneous images and constructing different copula mixtures to model their dependence structures based on image segmentation techniques [77]. It is also worth noting that this work is more of a feasibility study, whose significance is more of exploring the possibility of modeling the dependence structure between the heterogeneous RS images by means of copula mixtures as a first attempt.

Indeed, the components of the copula mixtures in this work are rather simple. In the future, we would utilize more complex copula functions to construct copula mixtures in order to attain comparable CD performance on more challenging large-scale RS datasets. We would also employ more sophisticated image translation techniques to enhance the quality of the translated images and to improve the CD performance of the proposed method.

## Acknowledgements

This work was supported by the National Key R&D Program of China under Grant 2021YFA0715201, the National Natural Science Foundation of China under Grant 61925106 and 62101303, and Autonomous Research Project of Department of Electronic Engineering at Tsinghua University.

## References

[1] V. Ferraris, N. Dobigeon, M. Chabert, Robust fusion algorithms for unsupervised change detection between multi-band optical images — A comprehensive case study, Inf. Fusion 64 (2020) 293-317.
[2] G. Cheng, G. Wang, J. Han, ISNet: Towards Improving Separability for Remote Sensing Image Change Detection, IEEE Trans. Geosci. Remote Sens. 60 (2022) 1-11.
[3] G. Xian, C. Homer, Updating the 2001 national land cover database impervious surface products to 2006 using Landsat imagery change detection methods, Remote Sens. Environ. 114 (8) (2010) 1676–1686.
[4] Z. Zhu, C. E. Woodcock, Continuous change detection and classification of land cover using all available landsat data, Remote Sens. Environ. 144 (2014) 152–171.
[5] M. Hussain, D. Chen, A. Cheng, H. Wei, D. Stanle, Change detection from remotely sensed images: From pixel-based to object-based approaches, J. Photogramm. Remote Sens. 80 (2013) 91-106.
[6] Y. Zhou, Y. Feng, S. Huo, X. Li, Joint Frequency-Spatial Domain Network for Remote Sensing Optical Image Change Detection, IEEE Trans. Geosci. Remote Sens. 60 (2022) 1-14.
[7] L. Ru, B. Du, C. Wu, Multi-Temporal Scene Classification and Scene Change Detection with Correlation Based Fusion, IEEE Trans. Image Process. 30 (2021) 1382-1394.
[8] W. Zhang, L. Jiao, F. Liu, S. Yang, J. Liu, Adaptive Contourlet Fusion Clustering for SAR Image Change Detection, IEEE Trans. Image Process. 31 (2022) 2295-2308.
[9] S. Marchesi, F. Bovolo, L. Bruzzone, A Context-Sensitive Technique Robust to Registration Noise for Change Detection in VHR Multispectral Images, IEEE Trans. Image Process. 19 (7) (2010) 1877-1889.
[10] K. Zhang, X. Lv, H. Chai, J. Yao, Unsupervised SAR Image Change Detection for Few Changed Area Based on Histogram Fitting Error Minimization, IEEE Trans. Geosci. Remote Sens. 60 (2022) 1-19.
[11] W. Fang, C. Xi, Land-Cover Change Detection for SAR Images Based on Biobjective Fuzzy Local Information Clustering Method With Decomposition, IEEE Geosci. Remote Sens. Lett. 19 (2022) 1-5.
[12] X. Jiang, G. Li, X. P. Zhang, Y. He, A semisupervised Siamese network for efficient change detection in heterogeneous remote sensing images, IEEE Trans. Geosci. Remote Sens. 60 (2021) 1-18.
[13] Y. Wu, J. Li, Y. Yuan, A. K. Qin, Q. -G. Miao, M. -G. Gong, Commonality Autoencoder: Learning Common Features for Change Detection From Heterogeneous Images, IEEE Trans. Neural Netw. Learn. Syst. 33 (9) (2022) 4257-4270.
[14] X. Li, Z. Du, Y. Huang, Z. Tan, A deep translation (GAN) based change detection network for optical and SAR remote sensing images, J. Photogramm. Remote Sens. 179 (2021) 14-34.
[15] Y. Sun, L. Lei, X. Li, H. Sun, G. Kuang, Nonlocal patch similarity-based heterogeneous remote sensing change detection, Pattern Recognit. 109 (2021) 107598.
[16] J. Prendes, M. Chabert, F. Pascal, A. Giros, J. Y. Tourneret, A new multivariate statistical model for change detection in images acquired by homogeneous and heterogeneous sensors, IEEE Trans. Image Process. 24 (3) (2015) 799–812.
[17] J. Liu, M. Gong, K. Qin, P. Zhang, A Deep Convolutional Coupling Network for Change Detection Based on Heterogeneous Optical and Radar Images, IEEE Trans. Neural Netw. Learn. Syst. 29 (3) (2018) 545-559.
[18] R. Touati, M. Mignotte, M. Dahmane, Anomaly feature learning for unsupervised change detection in heterogeneous images: A deep sparse residual model, IEEE J. Sel. Topics Appl. Earth Observ. Remote Sens. 13 (2020) 588–600.
[19] Z. -G. Liu, Z. -W. Zhang, Q. Pan, L. -B. Ning, Unsupervised Change Detection From Heterogeneous Data Based on Image Translation, IEEE Trans. Geosci. Remote Sens. 60 (2022) 1-13.
[20] Y. Sun, L. Lei, D. Guan, G. Kuang, Iterative Robust Graph for Unsupervised Change Detection of Heterogeneous Remote Sensing Images, IEEE Trans. Image Process. 30 (2021) 6277-6291.

[21] Y. Sun, L. Lei, X. Li, X. Tan, G. Kuang, Structure Consistency-Based Graph for Unsupervised Change Detection With Homogeneous and Heterogeneous Remote Sensing Images, IEEE Trans. Geosci. Remote Sens. 60 (2022) 1-21.
[22] G. Mercier, G. Moser, S. B. Serpico, Conditional copulas for change detection in heterogeneous remote sensing images, IEEE Trans. Geosci. Remote Sens. 46 (5) (2008) 1428–1441.
[23] P. J. Jaworski, F. Durante, W. K. Härdle, T. Rychlik, Copula Theory and Its Applications, Heidelberg, Germany, Springer, 2010.
[24] R. B. Nelsen, An Introduction to Copulas, New York, NY, USA, Springer, 2006.
[25] A. Thakkar, K. Chaudhari, Fusion in stock market prediction: A decade survey on the necessity, recent developments, and potential future directions, Inf. Fusion 65 (2021) 95-107.
[26] J. C. Rodriguez, Measuring financial contagion: A copula approach, J. Empir. Finance 14 (3) (2007) 401-423.
[27] X. Zhang, H. Jiang, Application of Copula function in financial risk analysis, Comput. Electr. Eng. 77 (2019) 376-388.
[28] F. Hawas, A. Cifuentes, Valuation of projects with minimum revenue guarantees: a gaussian copula-based simulation approach, Eng. Economist 62 (1) (2017) 90–102.
[29] E. N. Karimalis, N. K. Nomikos, Measuring systemic risk in the European banking sector: a copula CoVar approach, Eur. J. Finance 24 (1) (2018) 1–38.
[30] M. J. Lu, C. Y. H. Chen, W. K. Härdle, Copula-based factor model for credit risk analysis, Rev. Quant. Finance Account 49 (4) (2017) 1–23.
[31] A. Voisin, V. A. Krylov, G. Moser, S. B. Serpico, J. Zerubia, Supervised Classification of Multisensor and Multiresolution Remote Sensing Images With a Hierarchical Copula-Based Approach, IEEE Trans. Geosci. Remote Sens. 52 (6) (2014) 3346-3358.
[32] A. Voisin, V. A. Krylov, G. Moser, S. B. Serpico, J. Zerubia, Classification of Very High Resolution SAR Images of Urban Areas Using Copulas and Texture in a Hierarchical Markov Random Field Model, IEEE Geosci. Remote Sens. Lett. 10 (1) (2013) 96-100.
[33] V. A. Krylov, G. Moser, S. B. Serpico, J. Zerubia, Supervised HighResolution Dual-Polarization SAR Image Classification by Finite Mixtures and Copulas, IEEE J. Sel. Topics Signal Process. 5 (3) (2011) 554-566.
[34] L. Hu, Dependence patterns across financial markets: a mixed copula approach, Appl. Econ. 16 (10) (2006) 717-729.
[35] B. Y. Liu, Q. Ji, Y. Fan, Dynamic return-volatility dependence and risk measure of CoVaR in the oil market: A time-varying mixed copula model, Energ. Econ. 68 (2017) 53-65.
[36] E. Turgutlu, B. Ucer, Is global diversification rational? Evidence from emerging equity markets through mixed copula approach, Appl. Econ. 42 (5) (2010) 647-658.
[37] R. Dian, S. Li, B. Sun, A. Guo, Recent advances and new guidelines on hyperspectral and multispectral image fusion, Inf. Fusion 69 (2021) 40-51.
[38] P. Isola, J.-Y. Zhu, T. Zhou, A. A. Efros, Image-to-Image translation with conditional adversarial networks, in: Proc. IEEE Conf. Comput. Vis. Pattern Recognit. (CVPR), 2017, pp. 1125–1134.
[39] J.-Y. Zhu, T. Park, P. Isola, A. A. Efros, Unpaired Image-to-Image translation using cycle-consistent adversarial networks, in: Proc. IEEE Int. Conf. Comput. Vis. (ICCV), 2017, pp. 2223–2232.
[40] D. Zhu, X. Wang, G. Li, X-P. Zhang, Vessel detection via multi-order saliency-based fuzzy fusion of spaceborne and airborne SAR images, Inf. Fusion 89 (2023) 473-485.
[41] J. Wang, C. Tang, Z. Li, X. Liu, W. Zhang, E. Zhu, L. Wang, Hyperspectral band selection via region-aware latent features fusion based clustering, Inf. Fusion 79 (2022) 162-173.
[42] L. Khelifi, M. Mignotte, EFA-BMFM: A multi-criteria framework for the fusion of colour image segmentation, Inf. Fusion 38 (2017) 104-121.
[43] U. Wegmuller, C. Werner, Retrieval of vegetation parameters with SAR interferometry, IEEE Trans. Geosci. Remote Sens. 35 (1) (1997) 18-24.
[44] A. Stein, F. Van der Meer, B. Gorte, Spatial Statistics for Remote Sensing, Norwell, MA, USA: Kluwer, 1999.
[45] F. Biondi, A Polarimetric Extension of Low-Rank Plus Sparse Decomposition and Radon Transform for Ship Wake Detection in Synthetic Aperture Radar Images, IEEE Geosci. Remote Sens. Lett. 16 (1) (2019) 75-79.
[46] Y. -J. Liu, M. Yu, B. -J. Li, Y. He, Intrinsic manifold SLIC: A simple and efficient method for computing content-sensitive superpixels. IEEE Trans. Pattern Anal. Mach. Intell. 40 (3) (2018) 653-666.
[47] R. Achanta, A. Shaji, K. Smith, A. Lucchi, P. Fua, S. Süsstrunk, SLIC Superpixels Compared to State-of-the-Art Superpixel Methods, IEEE Trans. Pattern Anal. Mach. Intell. 34 (11) (2012) 2274-2282.
[48] Y. Ding, Z. Zhang, X. Zhao, Y. Cai, S. Li, B. Deng, W. Cai, Self-supervised locality preserving low-pass graph convolutional embedding for large-scale hyperspectral image clustering, IEEE Trans. Geosci. Remote Sens. 60 (2022) 1-16.
[49] E. C. Brechmann, Truncated and simplified regular vines and their applications, Diploma thesis, Technische Universität München, 2010.
[50] R. Liu, C. Yuen, T. N. Do, M. Zhang, Y. L. Guan, U. X. Tan, Cooperative positioning for emergency responders using self IMU and peer-to-peer radios measurements, Inf. Fusion 56 (2020) 93-102.
[51] M. G. Kendall, Rank Correlation Methods. Griffin, 1970.
[52] E. Bouyé, V. Durrleman, A. Nikeghbali, G. Riboulet, T. Roncalli, Copulas for Finance—A Reading Guide and Some Applications SSRN eLibrary, 2000 [Online]. Available: http://ssrn.com/paper=1032533.
[53] S. G. Iyengar, P. K. Varshney, T. Damarla, A Parametric CopulaBased Framework for Hypothesis Testing Using Heterogeneous Data, IEEE Trans. Signal Process. 59 (5) (2011) 2308-2319.
[54] S. K. Ng, T. Krishnan, G. J. McLachlan, The EM algorithm, in Handbook of computational statistics (pp. 139-172). Springer, Berlin, Heidelberg, 2012.

[55] T. Li, Z. Hu, Z. Liu, X. Wang, Multi-sensor Suboptimal Fusion Student's t Filter, IEEE Trans. Aerosp. Electron. Syst. (2022) early access.
[56] C. Yu, S. Li, D. Ghista, Z. Gao, H. Zhang, J. Del Ser, L. Xu, Multi-level multi-type self-generated knowledge fusion for cardiac ultrasound segmentation, Inf. Fusion 92 (2023) 1-12.
[57] X. Wang, D. Zhu, G. Li, X.-P. Zhang, Y. He, Proposal-Copula-Based Fusion of Spaceborne and Airborne SAR Images for Ship Target Detection, Inf. Fusion 77 (2022) 247-260.
[58] P. Du, S. Liu, J. Xia, Y. Zhao, Information fusion techniques for change detection from multi-temporal remote sensing images, Inf. Fusion 14 (1) (2013) 19-27.
[59] L.T. Luppino, M. Kampffmeye, F.M. Bianchi, G. Moser, S.B. Serpico, R. Jenssen, S.N. Anfinsen, Deep image translation with an affinity-based change prior for unsupervised multimodal change detection, IEEE Trans. Geosci. Remote Sens. 60 (2022) 1-22.
[60] X. Niu, M. Gong, T. Zhan, Y. Yang, A conditional adversarial network for change detection in heterogeneous images, IEEE Geosci. Remote Sens. Lett. 16 (2019) 45–49.
[61] L.T. Luppino, M.A. Hansen, M. Kampffmeyer, F.M. Bianchi, G. Moser, R. Jenssen, S.N. Anfinsen, Code-aligned autoencoders for unsupervised change detection in multimodal remote sensing images, IEEE Trans. Neural Netw. Learn. Syst. (2022) early access.
[62] X. Wang, W. Cheng, Y. Feng, R. Song, TSCNet: Topological Structure Coupling Network for Change Detection of Heterogeneous Remote Sensing Images, Remote Sens. 15 (3) (2023) 621.
[63] R. Touati, M. Mignotte, M. Dahmane, Multimodal change detection in remote sensing images using an unsupervised pixel pairwise-based Markov random field model, IEEE Trans. Image Process. 29 (2020) 757–767.
[64] Z.-G. Liu, G. Mercier, J. Dezert, Q. Pan, Change detection in heterogeneous remote sensing images based on multidimensional evidential reasoning, IEEE Geosci. Remote Sens. Lett. 11 (1) (2014) 168–172.
[65] L. Lei, Y. Sun, G. Kuang, Adaptive Local Structure Consistency-Based Heterogeneous Remote Sensing Change Detection, IEEE Geosci. Remote Sens. Lett. 19 (2022) 1-5.
[66] Y. Sun, L. Lei, X. Li, X. Tan, G. Kuang, Patch similarity graph matrix-based unsupervised remote sensing change detection with homogeneous and heterogeneous sensors, IEEE Trans. Geosci. Remote Sens. 59 (6) (2021) 4841–4861.
[67] Y. Sun, L. Lei, D. Guan, M. Li G. Kuang, Sparse-Constrained Adaptive Structure Consistency-Based Unsupervised Image Regression for Heterogeneous Remote-Sensing Change Detection, IEEE Trans. Geosci. Remote Sens. 60 (2022) 1-14.
[68] Q. Shi, M. Liu, S. Li, X. Liu, F. Wang, L. Zhang, A Deeply Supervised Attention Metric-Based Network and an Open Aerial Image Dataset for Remote Sensing Change Detection, IEEE Trans. Geosci. Remote Sens. 60 (2022) 1-16.
[69] Z. Lv, F. Wang, G. Cui, J. A. Benediktsson, T. Lei, W. Sun, Spatial–Spectral Attention Network Guided With Change Magnitude Image for Land Cover Change Detection Using Remote Sensing Images, IEEE Trans. Geosci. Remote Sens. 60 (2022) 1-12.
[70] H. Chen, Z. Qi, Z. Shi, Remote Sensing Image Change Detection With Transformers, IEEE Trans. Geosci. Remote Sens. 60 (2022) 1-14.
[71] H. Zhang, M. Lin, G. Yang, L. Zhang, ESCNet: An End-to-End Superpixel-Enhanced Change Detection Network for Very-High-Resolution Remote Sensing Images, IEEE Trans. Neural Netw. Learn. Syst. 34 (1) (2023) 28-42.
[72] V. Michele, C.-V. Gustau, T. Devis, Spectral alignment of multi-temporal cross-sensor images with automated kernel canonical correlation analysis, J. Photogramm. Remote Sens. 107 (2015) 50–63.
[73] R. Touati, M. Mignotte, M. Dahmane, Change Detection in Heterogeneous Remote Sensing Images Based on an Imaging Modality-Invariant MDS Representation, in: Proc. IEEE Int. Conf. Image Process. (ICIP), Athens, Greece, 2018, pp. 3998-4002.
[74] R. Shao, C. Du, H. Chen, J. Li, SUNet: Change Detection for Heterogeneous Remote Sensing Images from Satellite and UAV Using a Dual-Channel Fully Convolution Network, Remote Sens. 13(18) (2021) 3750.
[75] I. Bakkouri, K. Afdel, MLCA2F: Multi-level context attentional feature fusion for COVID-19 lesion segmentation from CT scans, Signal Image Video Process. (2022) 1–8.
[76] I. Bakkouri, K. Afdel, J. Benois-Pineau, et al. BG-3DM2F: Bidirectional gated 3D multi-scale feature fusion for Alzheimer's disease diagnosis, Multimed. Tools Appl. 81 (2022) 10743–10776.
[77] I. Chaturvedi, Q. Chen, R. E. Welsch, K. Thapa, E. Cambria, Gaussian correction for adversarial learning of boundaries, Signal Process. Image Commun. 109 (2022) 116841.